\documentclass[a4paper,12pt,twoside]{article} 
\usepackage[total={11in,8.5in}]{geometry}

 \usepackage[authoryear,round]{natbib}
\usepackage{multirow}
\usepackage{algorithm}
\usepackage[noend]{algpseudocode}
\usepackage{setspace}
\usepackage{lipsum}

 \usepackage[usenames]{color}
\usepackage[usenames,dvipsnames]{xcolor}
 \usepackage{textcomp}
\usepackage[latin1]{inputenc}  

\usepackage{color,soul}
\usepackage{float}
\usepackage{fancyvrb}
\usepackage{hyperref}
 \usepackage{verbatim}
\hypersetup{linkbordercolor=1 1
1,colorlinks=false,linkcolor=blue,pdfborder=0 0 0}
\usepackage{bbold}
\usepackage{xcolor}
\usepackage{lipsum}

\usepackage[catalan,spanish,english]{babel}
\usepackage{lscape}
\usepackage{latexsym, graphicx, graphics}
\usepackage{amsmath}
\usepackage{mathtools}
\usepackage[usenames]{color}	
\usepackage{tikz}
\usepackage{amssymb}
\usepackage[labelfont=bf,labelsep=period,justification=raggedright]{caption}
\usepackage{tabularx}
\usepackage{subfig}
\usepackage{graphicx}
\makeatletter
\def\BState{\State\hskip-\ALG@thistlm}
\makeatother

\def\dg{g}
\def\dc{d}

\let\emptyset\varnothing

\renewenvironment{abstract}
 {\footnotesize
  \begin{center}
  \bfseries \abstractname\vspace{-.7em}\vspace{0pt}
  \end{center}
  \list{}{
    \setlength{\leftmargin}{.7cm}%
    \setlength{\rightmargin}{\leftmargin}%
  }%
  \item\relax}
 {\endlist}

\begin{document}

\graphicspath{{./GRAPHICS/}}

\pagenumbering{arabic}
\selectlanguage{english}

\pagestyle{plain}

{\Large
\vspace*{0.5cm}
\textbf{Selection of the Regularization Parameter
in\\
 Graphical Models using Network Characteristics\\
}}
\vspace{0.2cm}
\\
\vspace{0.1cm} Adria Caballe Mestres$^{a,b}$, Natalia Bochkina$^a$ and Claus Mayer$^b$\\
$^a$ University of Edinburgh \& Maxwell Institute, Scotland, UK \\
$^b$ Biomathematics \& Statistics Scotland, Scotland, UK

\begin{abstract}
\onehalfspacing
\footnotesize
Gaussian graphical models represent the underlying graph structure of conditional dependence between random variables which can be determined using their partial correlation or precision matrix. In a high-dimensional setting, the precision matrix is estimated using penalized likelihood by adding a penalization term which controls the amount of sparsity in the precision matrix and totally characterizes the complexity and structure of the graph. The most commonly used penalization term is the L1 norm of the precision matrix scaled by the regularization parameter which determines the trade-off between sparsity of the graph and fit to the data. In this paper we propose several procedures to select the regularization parameter in the estimation of graphical models that focus on recovering reliably the appropriate network structure of the graph. We conduct an extensive simulation study to show that the proposed methods produce useful results for different network topologies. The approaches are also applied in a high-dimensional real case study of gene expression data with the aim to discover the genes relevant to colon cancer. Using this data, we find graph structures which are verified to display significant biological gene associations. 
	\end{abstract}

\onehalfspacing
\vspace{-0.8cm}
\onehalfspacing
\section{Introduction}\label{SEC1}
In recent years, the study of undirected graphical models \citep{Lauritzen2011} has been the focus of attention of many authors. The increasing volume of high-dimensional data in different disciplines makes them a useful tool in order to determine conditional dependence between random variables. For instance, graphical models have been applied to gene expression data sets to find biological associations across genes in \cite{Dobra2004} and \cite{Schafer2005}, as well as in other biological networks \citep{Newman} and  in social networks \citep{Goldenberg2007}. In Gaussian graphical models, which are often used for finding associations between genes using high throughput genomic data, the dependence between the genes is fully characterized by the non-zero elements of the precision matrix $\Omega$ (see Section~\ref{SEC:model}).

However, in a high dimensional framework where the number of variables $p$ is larger than the number of observations $n$, there is not enough information in the data available to estimate $\Omega$, and hence the underlying conditional dependence (CD) graph. To address this problem, alternative estimators have been proposed in the last two decades using additional information about $\Omega$ such that the estimated covariance matrix and its inverse are of full rank. Typically, three classes of estimators of $\Omega$ have been used: thresholding \citep{Bickel2008}, shrinkage \citep{Ledoit2004, Daniels2001} and penalized log likelihood \citep{Tibshirani1996}. 
 
In this paper we consider the latter kind of estimators, the graphical Lasso penalization method (defined in Section \ref{SEC:model}) which adds the penalty $\lambda ||\Omega||_1$ with a tuning parameter $\lambda$ in the maximum likelihood. The penalized maximum likelihood optimization problem is solved using recursive algorithms, for instance we find that three of the most efficient and commonly used ways to solve it are GLasso by \cite{Friedman2007}, Neighborhood selection by \cite{Meinshausen2006} and Tuning-Insensitive Graph Estimation and Regression by \cite{Liu2012}.  The choice of the tuning parameter $\lambda$ represents the trade-off between close fit to the data and sparsity of $\Omega$, and its selection for estimation of the corresponding CD graph structure is the topic of this paper.  

Methods such as Cross Validation (CV), Akaike Information Criterion (AIC) or Bayesian Information Criterion (BIC) have been widely used to select tuning parameters when $p$ is small. However, they fail once dealing with high-dimensional problems by over-fitting the graph structure of $\Omega$ \citep{Liu2011, Wasserman2009}. 

\cite{Liu2011} proposed the selection of $\lambda$ by controlling the desirable approximated variability in the estimated graphs using a subsampling approach (StARS). This method contrasts with the usual variable selection statistics since it only considers the estimated CD graph structure. Even though the method is promising and gives an alternative to AIC and BIC, it has a major drawback:  another tuning parameter is needed in order to set the maximum variability across samples which can be unknown a priori in many applications. Our simulations show that the default values can lead to overestimation of the network size in certain graph topologies. \cite{Meinshausen2010} presented a stability selection approach which controls the graph edges false discovery rate. The authors estimate $\Omega$ by an average subsampling graphical Lasso method such that the effect of the choice of $\lambda$ is very low. However, the trade-off between false positive and true positive edges of the selected network by their subsampling approach is worse than the one given by a network with the same number of edges using all the data due to considering smaller effective sample sizes than the original $n$ for estimation. To the best of our knowledge, there is no other relevant approach in the literature that only uses the graph structure to select the tuning parameter $\lambda$ in graphical models.

We have applied the following methods for selecting $\lambda$ popular in statistical literature to estimate CD graph structures in microarray data: AIC, BIC and StARS. However, the graphs we have obtained were rather dense and very difficult to interpret to a biologist,  namely to extract groups of genes acting together and possibly interacting. In the biological literature, the most commonly used approaches to construct gene networks are based on clustering. This is informed by the expected presence of distinct strongly interconnected clusters in biological networks \citep{Eisen1998,Yi2007}.
This gave us the motivation to find $\lambda$ such that the  corresponding graph has  a clustering structure which can be interpreted by a biologist without  restricting it to a block diagonal structure and hence missing potentially important interactions.


Our aim is to select the hyperparameter $\lambda$ such that (a) it produces reliable estimates of the edges of the graph  (b) the corresponding CD graph structure is interpretable in terms of network characteristics and (c) works well for networks that arise in biological systems. In this paper, we propose several such approaches to selecting $\lambda$, in the framework of a general two-step procedure. The main novelty with respect to classical approaches such as AIC or BIC is that we use only the graph structure of the GLasso estimator to tune the regularization parameter $\lambda$. The first proposed approach, Path connectivity (PC), uses the average geodesic distance of estimated networks to find the graph that corresponds to the biggest change of the number of connections and is associated with splitting of clusters. The second method, Augmented mean square error (A-MSE), similarly to the StARS approach, controls the variability of the estimated networks in terms of graph dissimilarity coefficients using subsampling. The main difference from StARS is the additional bias term to avoid having a tuning parameter. We consider the bias with respect to an initial estimated graph structure which contains a desirable global network characteristic. For instance, we use the AGNES hierarchical clustering coefficient \citep{Kaufman2009}, which is the third proposed method to choose $\lambda$, to select the graph that presents the highest clustering structure. Although clustering methods exist in the literature, the novelty here is that we use them to select the penalty parameter $\lambda$ in Graphical Lasso estimation.

We compare performance of the proposed approaches as well as of the StARS algorithm and of the standard AIC and BIC on both simulated and real data. The data is a microarray gene expression data set generated by the TCGA Research Network: \url{http://cancergenome.nih.gov/}. It contains 154 samples for patients with colon tumor and about 18k genes. We are particularly interested in finding significant complex gene interactions reliably and relating the observed associations to  pathway databases which describe known biochemistry connections between genes. Simulations and real data analysis are performed using the free statistical software R \citep{Rref}.

The rest of the paper is organized as follows. In Section \ref{SEC2} we introduce the tuning parameter selection methodology and in Section \ref{SEC3} we give their main algorithmic and computational information. In Section \ref{SEC4} we compare the performance of the methods using simulated data and then apply them to a gene expression dataset in Section \ref{SEC5}.


\section{Regularization parameter selection}\label{SEC2}

\subsection{Gaussian graphical model}\label{SEC:model} 

We assume that the data are iid observations from a Gaussian model: $X_i \sim N_p(0,\Omega^{-1})$, $i=1,\ldots, n$ independently, assuming, without a loss of generality, that the mean is zero. 
Conditional dependence is totally characterized by the inverse covariance matrix $\Omega$, also called the precision matrix. 
Two  Gaussian random variables $X_i$ and $X_j$ are said to be conditionally independent given all the remaining variables if the coefficient $\Omega_{i,j}$ is zero. This is often expressed with a graph structure $G$ in which each node represents a random variable and there is an edge connecting two different nodes if the correspondent element in the inverse covariance matrix is non-zero.

The corresponding log likelihood function for $\Omega$ is $\ell(\Omega) = \log\det\Omega - tr(S\Omega)$ where $S = n^{-1} \sum_{i=1}^n X_i^2$. If $S^{-1}$ exists ($p<n$ is a necessary condition), the MLE of $\Omega$ is given by $S^{-1}$.  However, in a high dimensional framework where the number of variables $p$ is larger than the number of observations $n$, the matrix $S$ is singular and so cannot be inverted.

We make an additional assumption that the CD (conditional dependence) graph is sparse, and hence that the precision matrix $\Omega$ is sparse. 
 Ideally, we would like to use a penalized likelihood estimator, with the penalty proportional to the number of non-zero elements in $\Omega$. However, such   optimization problem is non-convex and thus is very computationally intensive. In practice, a likelihood estimator with a convex penalty term proportional to the $\ell_1$ norm of $\Omega$, a Graphical Lasso, is commonly used instead:
\begin{equation}\label{eq:ML2}
\hat{\Omega}_{PML}^{\lambda} = \arg\max\limits_{\Omega \succ 0}\mbox{ } [\log\det\Omega - tr(S\Omega) - \lambda ||\Omega||_1],
\end{equation}
where $||\Omega||_1 = \sum_{i,j=1}^p |\Omega_{ij}|$ is the element-wise $\ell_1$ norm of the matrix $\Omega$ and PML stands for penalized maximum likelihood. For small $\lambda$, the corresponding penalized estimator of $ \Omega$ tends to be dense and in the extreme ($\lambda=0$) we are back to the initial Maximum Likelihood problem which may not have unique solution when $p/n$ is large \citep{Pourahmadi2011}. As we increase $\lambda$, the matrix becomes more and more sparse until we get a diagonal matrix.  Therefore, the choice of  $\lambda$ has a crucial effect on the estimated CD graph structure.


\subsection{General two step procedure to select the tuning parameter}

The $\ell_1$ penalized maximum likelihood estimator defined in \eqref{eq:ML2} requires  selection of a regularization parameter $\lambda$.
If the $\ell_1$ penalization genuinely represented our true prior knowledge about $\Omega$ then one of the standard methods such as the maximum marginal likelihood or cross validation for the elements of $\Omega$ could be used. However, the $\ell_1$ penalty here is used due to its computational convenience, replacing the $\ell_0$ penalty, so these methods are not appropriate. It is well known for the problem of estimating sparse vectors in high dimensions with the Lasso penalty, that the variable selection part, with an appropriate $\lambda$, is consistent, however, the estimation of the non-zero values usually has some bias \citep{Wasserman2009,Gu2013}. This can be due to the convex relaxation of the desired $\ell_0$ penalty to the computationally efficient $\ell_1$ penalty. Therefore, in this paper we propose to employ the methods that use only the variable selection part from the GLasso, $\widehat{G}^\lambda$, for tuning the hyperparameter $\lambda$.

We propose the following two step procedure for estimating $\lambda$:
\begin{enumerate}
\item Set $\hat{\Omega}_{PML}^\lambda$ as in equation (\ref{eq:ML2}) for all $\lambda \in \Lambda$,  $\Lambda \subset [0,\lambda_{max}]$, $\lambda_{max}>0$.
\item Choose $\hat\lambda = \arg \min_{\lambda} R(\lambda, \widehat{G}^\lambda)$
\end{enumerate}
using risk functions $R$ that are based only on CD graphs $\widehat{G}^\lambda$. 
This procedure combines computational efficiency of the Lasso algorithm with the choice of $\lambda$ that optimizes relevant characteristics of the CD graph such as connectivity, clustering structure, etc.


\subsection{Graph notation and distances}\label{SEC21}

Before introducing the risk functions, we give some basic definitions and properties of networks \citep{daCosta2007,Estrada2011} which will be used to select the regularization parameter.

A graph $G(V,E)$ is a set of nodes $V$, with connections between them, called edges $E$. The graph structure is often represented by a $p\times p$ matrix, called adjacency matrix and denoted by $A_G$. In the estimation of graphical models, the off-diagonal elements of $A_G$ are determined by the precision matrix (0 if $\Omega_{ij} = 0$ and 1 otherwise) and the diagonal elements are set to zero. Note that graphical models are undirected which means that the correspondent $A_G$ is always symmetric.

The distance between a pair of nodes $V_i$ and $V_j$ $\in$  $G(V,E)$ (also known as the geodesic distance) defines the shortest number of edges connecting node $V_i$ to the node $V_j$ which is denoted by $\dg_{ij}$. If there is no path linking the two nodes, then $\dg_{ij} = \infty$.
The correlation coefficient $\sigma_{ij}$ between two nodes $V_i$, $V_j$ $\in$  $G(V,E)$ and the corresponding dissimilarity measure $\dc_{ij}$ are given by
\begin{equation}\label{eq:AG1}
\sigma_{ij} = \eta_{ij}/\sqrt{\kappa_i\kappa_j}, \mbox{ \hspace{0.3cm} with \hspace{0.3cm} } \dc_{ij} = 1-\sigma_{ij}, \mbox{ \hspace{0.3cm}} D  = [\dc_{ij}]
\end{equation}
where $\eta_{ij}$ is the number of neighbors shared by the nodes $V_i$ and $V_j$ and $\kappa_i$ is the degree of the node $V_i$ defined as the number of nodes that are directly connected to $V_i$. 

\subsection{Proposed risk functions}
 We propose several risk functions to select $\lambda$ that monitor network characteristics of the conditional dependence graphs 
 that can be applicable to genomic data.  It has been observed \citep{Yi2007} that molecules in a cell work together in groups, with some -- usually less strong -- interaction between the groups. This motivates our choice of risk functions to encourage a clustering structure in the estimated graphs.


\subsubsection{Path connectivity risk function}

To motivate the first proposed risk function, we observe the following obvious property of the  graph $\hat{G}^\lambda$ that corresponds to the penalized estimator $\hat{\Omega}^{\lambda}$ defined by \eqref{eq:ML2}: for small $\lambda$, the likelihood term dominates and the estimator $\widehat{G}^\lambda$ is usually a dense graph with $\hat\Omega^{\lambda}$ closely fitting the data, and for large $\lambda$, the penalty term dominates and the corresponding estimate is a very sparse graph with $\hat\Omega^{\lambda}$  not fitting the data well. Thus, for growing values of $\lambda$, there is a decrease in graph complexity, and the aim of the method we propose here is to capture the value of $\lambda$ that corresponds to the largest change in the complexity of the graph.

For simplicity, we consider a grid of values of $\lambda$, $\Lambda = (\lambda_k)_{k=1}^M$ such that $\lambda_k - \lambda_{k-1} = h$, $k=2, \ldots, M$, and  the underlying estimated graphs $\widehat{G}^\lambda$  for all $\lambda \in \Lambda$. 
%
 We propose Path connectivity (PC) which is a novel approach to find $\lambda$ that finds the biggest change in graph complexity between the graphs $\hat{G}^{\lambda}$ corresponding to two consecutive values of  $\lambda \in \Lambda$. In this case the measure of graph complexity is calculated by the \emph{geodesic distance mean} statistic
\begin{equation}\label{eq:PC1}
H(\lambda)= \frac{2}{p\,(p-1)} \sum\limits_{i<j} \hat{\dg}_{ij}(\lambda)  I(\hat{\dg}_{ij}(\lambda) < \infty),
\end{equation}
where $\hat{\dg}_{ij}(\lambda)$ are the dissimilarity coefficients for the graph $\hat{G}^{\lambda}$.
To find the largest change in $H(\lambda)$, we consider the first order differences of $H(\lambda)$  by $D_h(\lambda) = \Delta_h H(\lambda)$, where $\Delta_{h}$ refers to the difference operator with bandwidth $h$. The regularization parameter selection by PC is given by the $\lambda$ that produces the most rapid relative descent in the number of graph connections:
\begin{equation}\label{eq:PC2}
\lambda_{pc} = \arg\max\limits_{\lambda_k \in \Lambda} R_{PC}(\lambda_k) = \arg\max\limits_{\lambda_k \in \Lambda} \left|D_{h}(\lambda_k)/\bar{D}_h(\lambda_k)\right|,
\end{equation}
where $\lambda_k$ is the $k$-th ordered element in $\Lambda$ and $\bar{D}_h(\lambda_k)$  is the running average defined as the average of elements $D_h(\lambda)$ with $\lambda \in \{ \lambda_1, \ldots ,  \lambda_k\}$. The difference of the geodesic distance mean is divided by $\bar{D}_h(\lambda_k)$ in ~\eqref{eq:PC2}  to favor  big jumps for larger $\lambda_k$ (and sparser $\hat{G}^\lambda$)  in comparison to the jumps for smaller $\lambda_k$ which correspond to more dense graphs.



\begin{figure}[ht]
\begin{center}
 \begin{tabular}{ccc}
   \subfloat[True clustered network]{\includegraphics[scale=0.24]{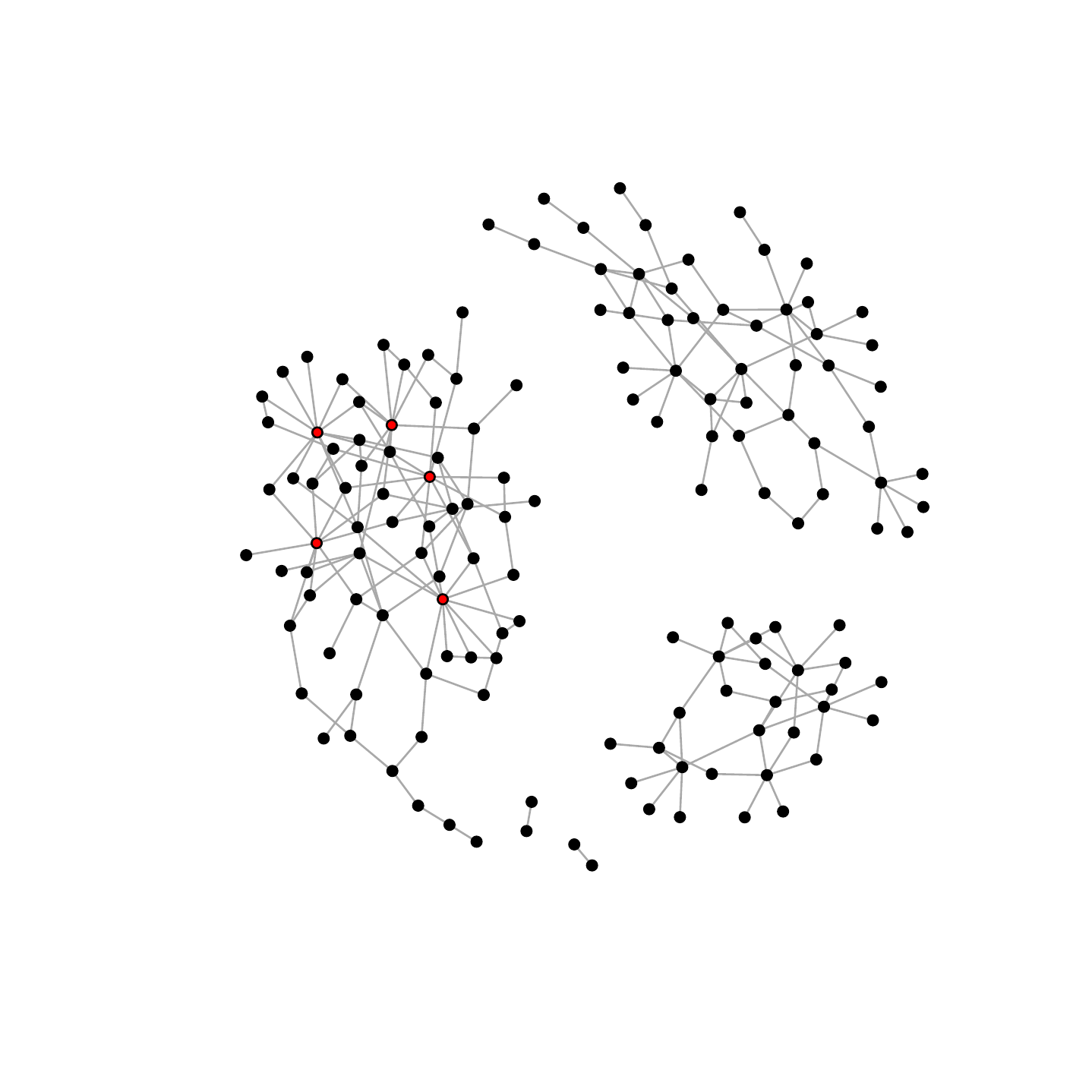}}&
\subfloat[Est. graph with $\lambda=\lambda_{pc}$ ]{\includegraphics[scale=0.24]{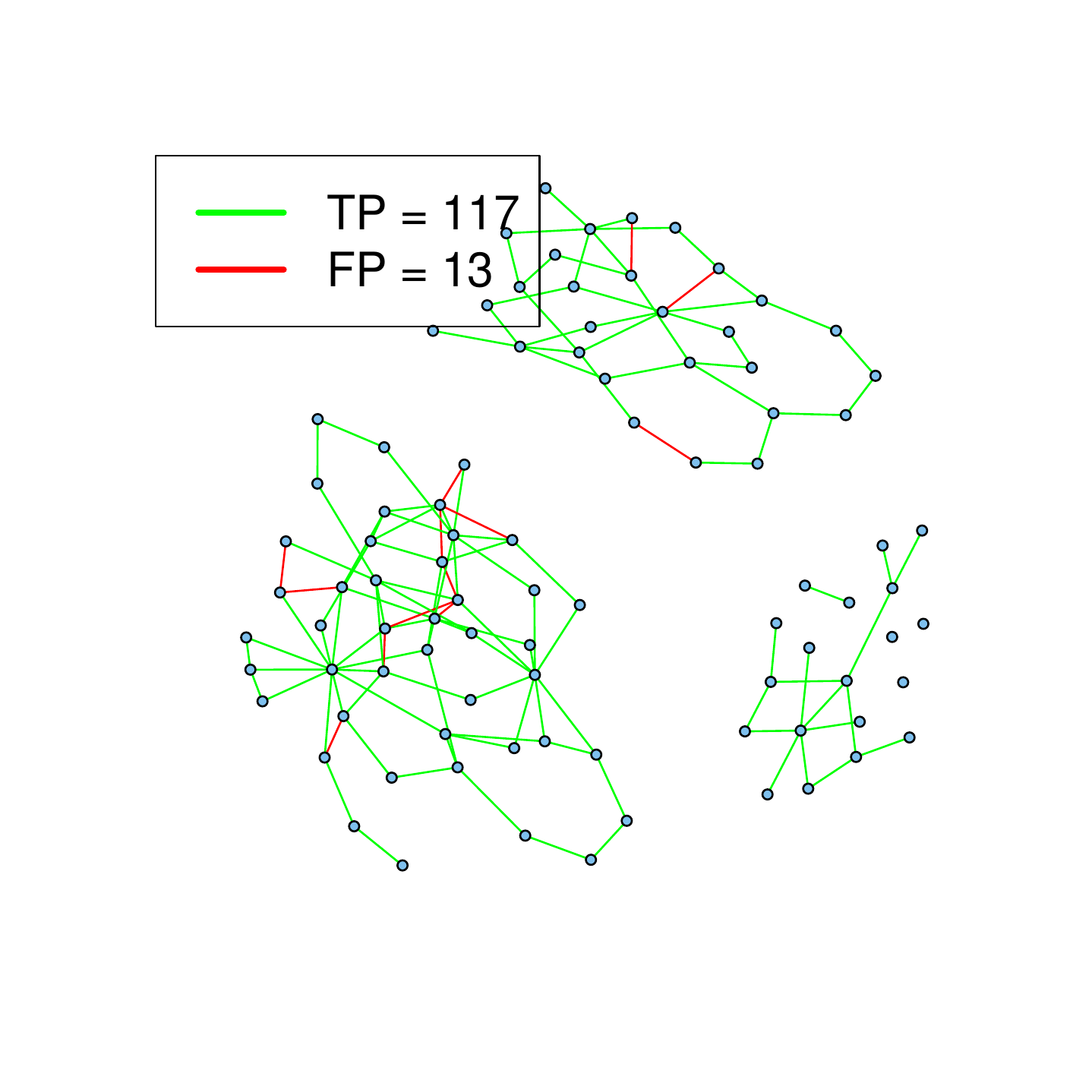}}
   & \subfloat[Est. graph with $\lambda=\lambda_{pc}-1$.]{\includegraphics[scale=0.24]{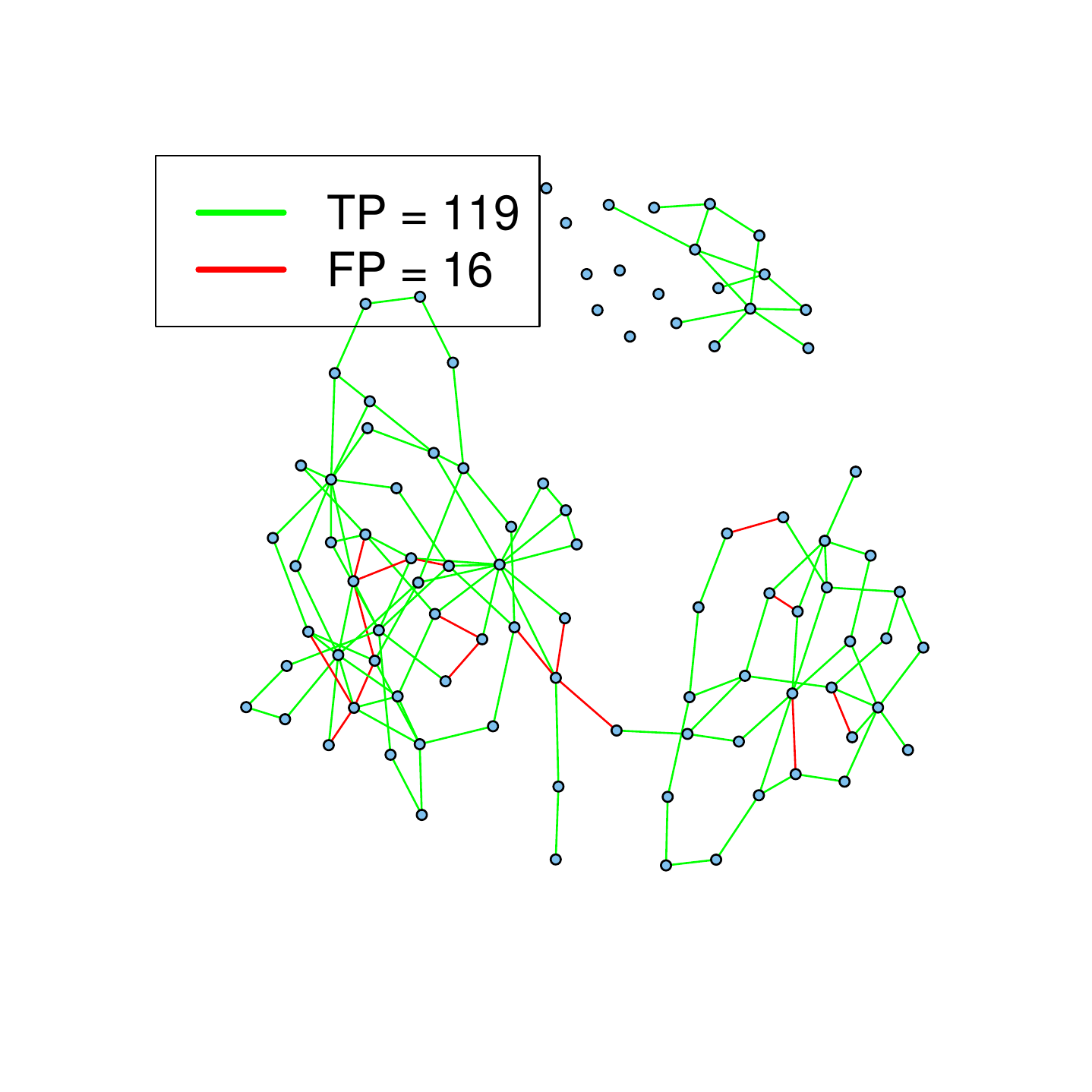}}\\
\subfloat[Geodesic distance mean ]{\includegraphics[scale=0.24]{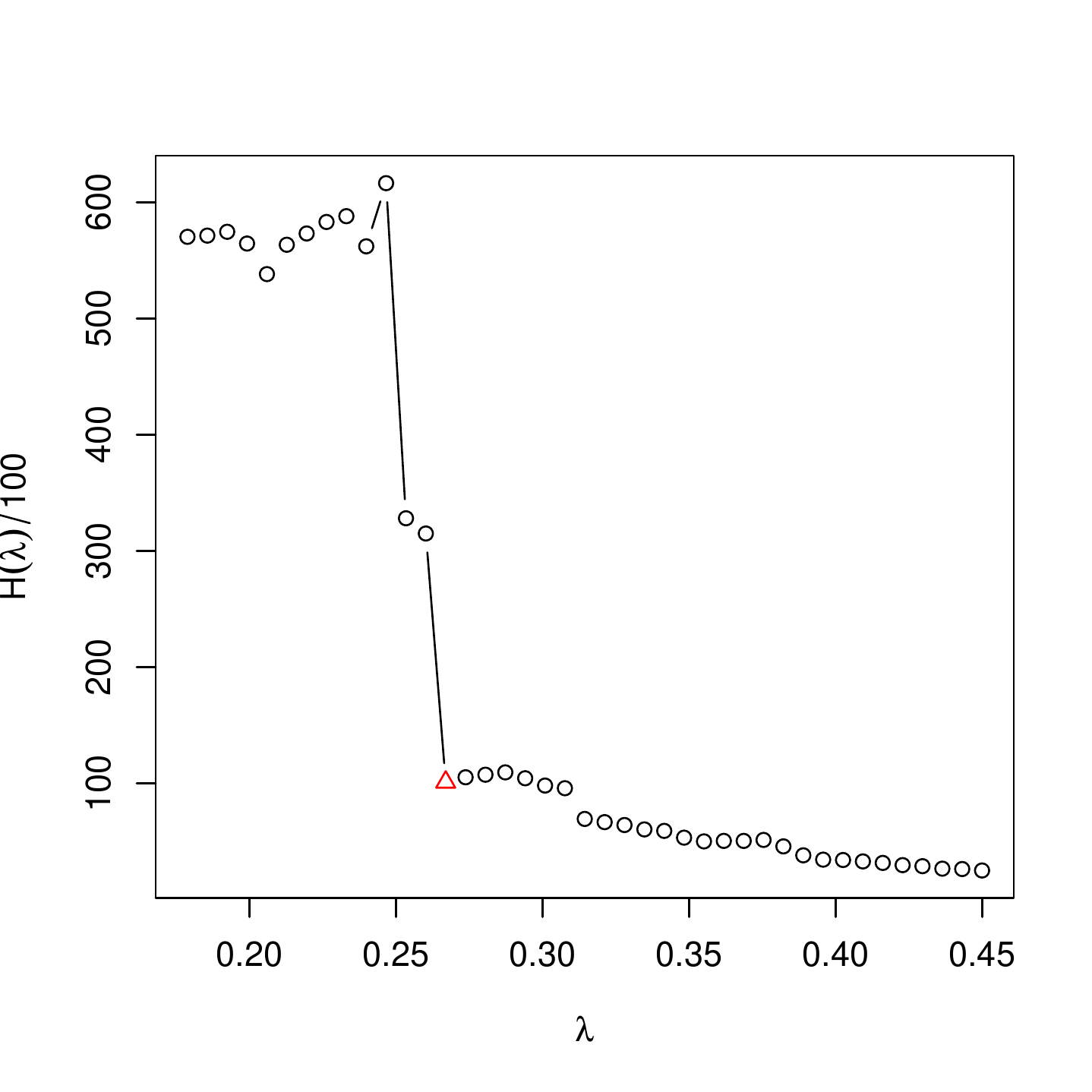}}&
\subfloat[Densities of $\hat{\lambda}=\lambda_{pc}$]{\includegraphics[scale=0.24]{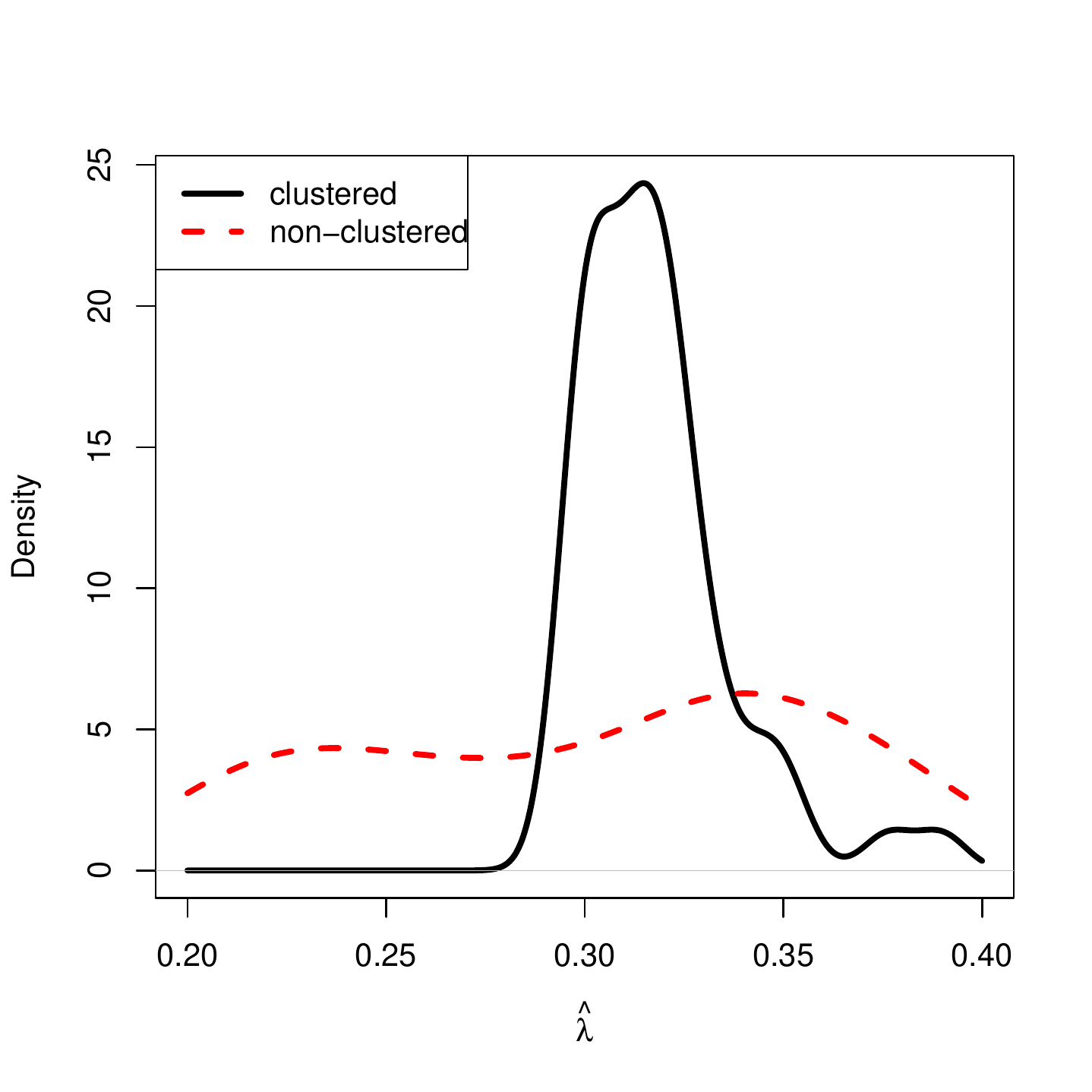}}&
\subfloat[True non-clustered network]{\includegraphics[scale=0.24]{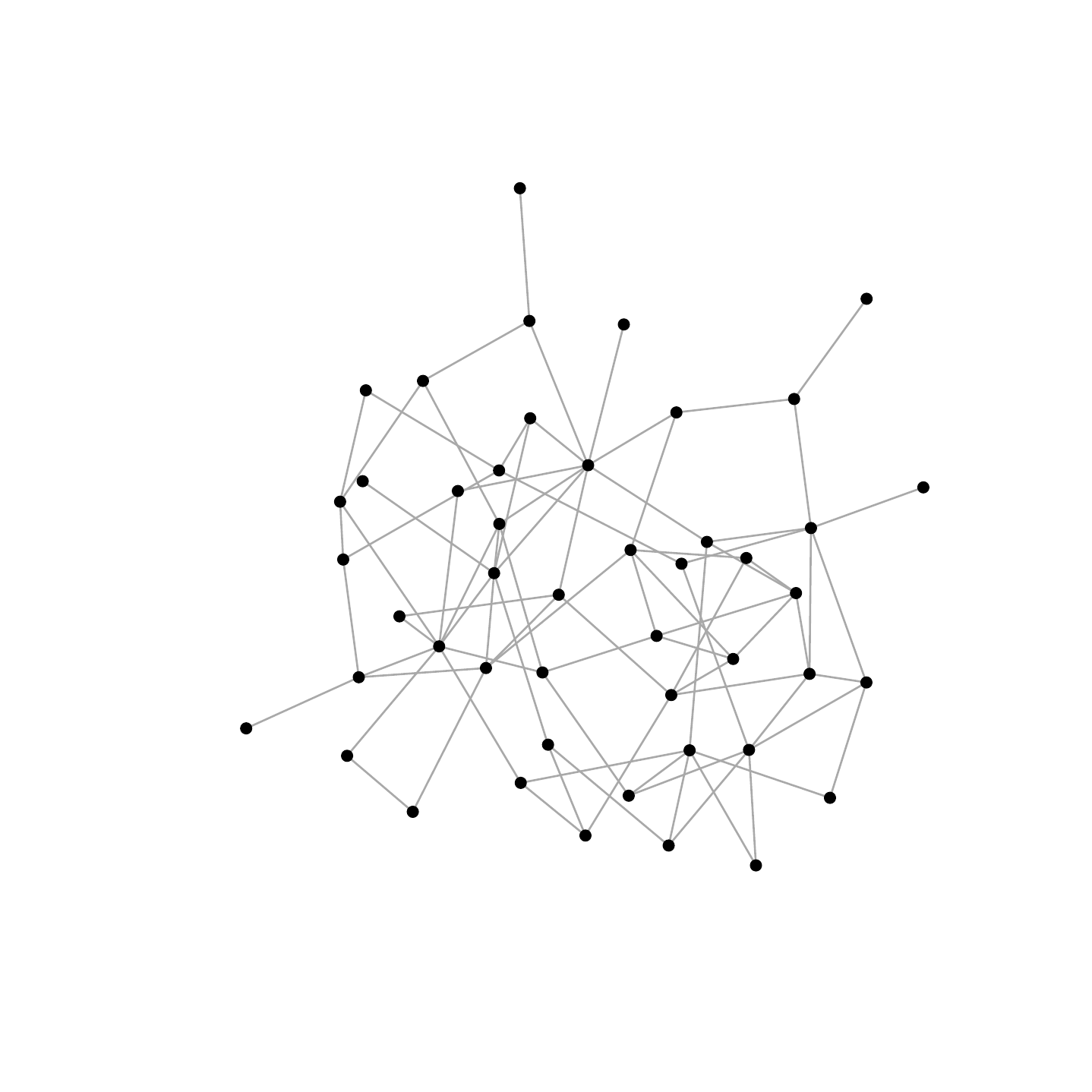}}
\end{tabular}
 \caption{{\footnotesize Path connectivity regularization parameter selection using a clustered -true network in (a) and solid line in (e)- and a non-clustered  -true network in (f) and dashed line in (e)- graph structures.}}\label{f8}
\end{center}
\end{figure}

In Figure \ref{f8} we illustrate the motivation of using the PC selection of $\lambda$ in simulated data (see Section \ref{SEC4} for details). The true CD graph structure defined by three non-overlapping clusters is plotted in Figure  \ref{f8}(a). We show the  geodesic distance mean as function of $\lambda$ for graph estimations in Figure  \ref{f8}(d).
It shows two big jumps which are related to the separation of clusters. The latter one gives the selected graph by PC and is due to the separation of two clusters (see Figure  \ref{f8}(b) for the selected $\lambda_{pc} = \lambda_k$ and Figure  \ref{f8}(c) for the previous graph structure defined by $\lambda_{k-1}$). This is a generally observed behaviour in both simulated and real gene expression datasets. In Figure  \ref{f8}(e) we show the density estimates of $\lambda_{pc}$ using 100 iid datasets with $n=100$, $p=170$ and two theoretical graph structures: hubs-based clustered graph as shown in Figure  \ref{f8}(a) and  non-clustered/random graph structure as shown in Figure  \ref{f8}(f). We can see the clear peak around $\lambda =0.33$ for the clustered data against a relatively flat empirical distribution for the non-clustered data.

\subsubsection{A-MSE risk function}


The idea explored in this section is to use a risk function based on network characteristics such as dissimilarities of the graph defined by  \eqref{eq:AG1}. Ideally, we would like to find $\lambda^\text{oracle}$ that minimizes
\begin{equation}\label{eq:AAG1}
R_{MSE}(\lambda) = \mathbf{E}(\sum_{i>j} |\dc_{ij}-\hat{\dc}_{ij}(\lambda)|^q), \mbox{ \hspace{0.3cm}}
\end{equation}
for some $q\geq 1$ where $\dc_{ij}$ are the dissimilarities of the true graph defined by \eqref{eq:AG1} and $\hat{\dc}_{ij}(\lambda)$ are the dissimilarities of the CD graph estimated by \eqref{eq:ML2} for a given tuning parameter $\lambda$. $R_{MSE}(\lambda)$ depends on the unknown true graph structure of $\Omega$; in practice, an unbiased estimator of $R_{MSE}(\lambda)$ is used, commonly obtained by subsampling (bootstrap, cross validation) by comparing estimated values to observations.  However, the problem in this setting is that direct observations of $d_{ij}$ are not available.

To overcome this problems we propose to use an initial graph estimate $\tilde{G}$ and its dissimilarities coefficients $[\tilde{d}_{ij}]$ in place of observed data. Thus, we propose to use the following choice of $\lambda$:


\begin{equation}\label{eq:AAGq}
\lambda_{amse} = \arg\min_{\lambda \in \Lambda} \hat{R}_{AMSE}(\lambda) = \arg\min_{\lambda \in \Lambda} \sum_{i>j}  \hat{\mathbf{E}}| \tilde{\dc}_{ij} -\hat{\dc}_{ij}(\lambda)|^q,
\end{equation}
where $\hat{\mathbf{E}}$ indicates the estimation of the expected value using subsampling. The algorithm is given in Section ~\ref{AAG}.  We find that $\lambda_{amse}$ can approximate well $\lambda^{oracle}$ in our simulated data (see Section 5 in Supplementary material).

For $q=2$, this risk function can be written as a sum of the variance terms and the sum of the squared differences between the initial and the current estimator (the ``bias'' term); see equation \eqref{eq:AAG2} in Section ~\ref{AAG}.
  Note that the first summand in (\ref{eq:AAG2}), the variance of the estimated distances, gives a stability measure similar to the one proposed in StARS (the latter uses the adjacency matrix instead of the dissimilarities). However, we add a bias term for the distance estimator which allows us to avoid the selection of the power tuning parameter $\beta$ that controls the desired variability in the StARS approach \citep{Liu2011}.


The proposed $R_{AMSE}(\lambda)$ risk can be applied to other network characteristics. By the definition of graph dissimilarities, $d_{ij}=1$ if nodes $i$ and $j$ are neither directly nor indirectly (share neighbor) connected. Defining $h_{ij}=0$ if $\sigma_{ij} = 1- d_{ij} = 0$ and $h_{ij}=1$ if $\sigma_{ij}>0$,  for sparse networks, there are many $h_{ij} = 0$ and only few $h_{ij} = 1$. Applying the $R_{AMSE}(\lambda)$ to $[h_{ij}]$ instead of $[d_{ij}]$,  we obtain
$$
R_{AMSE}^h(\lambda) = \mathbf{E} \sum_{i<j} (h_{ij} -\hat{h}_{ij}(\lambda))^2 = C_h + \mathbf{E} \sum_{(ij)\in \theta(\lambda)} (1-2 h_{ij}) = C_h + \mathbf{E} [TP(\lambda)-FP(\lambda)]
$$
where $\theta(\lambda) = \{(i,j); \, i<j \, \& \, \hat{h}_{ij}(\lambda) =0 \}$, FP$(\lambda)=\sum_{i<j}  I[h_{ij}=0, \,\hat{h}_{ij}(\lambda)=1]$, TP$(\lambda)=\sum_{i<j}  I[h_{ij}=1 , \, \hat{h}_{ij}(\lambda)=1]$ and $C_h$ is independent of $\lambda$. . Minimizing $R_{AMSE}^h(\lambda)$ is the same as maximizing the TP and FP differences (also known as Youden indices).

In practice, biologists often use clustering algorithms to discover groups of genes. Hence, we propose to use the output of a hierarchical clustering algorithm as an initial estimate of the graph to characterize global structure for the dissimilarities $[d_{ij}]$. 
We have investigated several clustering algorithms on real and simulated data, and we have not found much difference in the resulting graph estimate.
Below we present the algorithm based on AGNES clustering method.



\subsubsection{AGNES risk function}

Clustering of features using a dissimilarity measure has been intensively studied in the literature. Here we focus on the algorithm AGNES (AGglomerative NESting) which is presented in \citet[chap. 5]{Kaufman2009} and is implemented in the R package \texttt{cluster} \citep{Rousseeuw2013}. AGNES finds clusters iteratively joining groups of nodes with the smallest average dissimilarity coefficient. This average is found by considering the dissimilarity coefficients between all possible pairs of nodes from two different clusters. Moreover, AGNES proposes an agglomerative coefficient (AC) that measures the average distance between a node in the graph and its closest cluster of nodes.  We propose to choose $\lambda$ that maximizes the AC coefficient
\begin{equation}\label{eq:AG3}
\lambda_{ac} = \arg\max\limits_{\lambda \in \Lambda} \hat{R}_{AGNES}(\lambda) =  \arg\max\limits_{\lambda \in \Lambda} AC(\lambda).
\end{equation}
The details of the AGNES algorithm and the definition of the coefficient AC can be found in Section \ref{SEC3}.

The matrix of dissimilarities $D$ obtained by (\ref{eq:AG1}) gives a good representation of the complexity of a given graph, so, in addition to being  applied as an initial estimate for the A-MSE method described above, AGNES can also be used as a method of choosing $\lambda$.

\subsection{Comparison of the methods}\label{sec:Comparison}

In Table \ref{tabPOWER9} we give some of the main properties of the 6 risk functions we want to compare which are the three proposed methods, as well as StARS \citep{Liu2011}, AIC and BIC. Likelihood-based risk functions to select $\lambda$ such as AIC and BIC are useful to compromise between goodness of fit to the data and model over-fitting. The additional AIC penalty (given by $p\,(p-1)$) is smaller than BIC (given by $p\,(p-1)\log(n)/2$) even for very small $n$. Hence, the selection of $\lambda$ by AIC results in a denser CD graph structure of $\Omega$ than by BIC. 
StARS gives a good alternative to select $\lambda$ when estimating graph structures. It transforms the selection of $\lambda$ problem to the choice of the maximum expected variability that we allow in the graph. Even though such a choice is more intuitive than the direct selection of $\lambda$, we find it difficult to use without any prior information; our simulations show that using the default value of the tuning parameter results in high number of false positive edges (see Section \ref{ROCc}).

We provide two computationally fast approaches, AGNES and PC, and the slightly more computationally challenging  A-MSE method due to subsampling. The AGNES selection tends to find the most clustered graph possible such that different groups of nodes can be interpreted and analyzed.  This is found to be a good choice of $\lambda$  to recover global graph structure characteristics when the true precision is block diagonal (See Section 4 in the supplementary material). 
The A-MSE selection uses the AGNES estimator as the initial graph structure with the aim to improve estimations of local network characteristics. The value of $\lambda$ selected by A-MSE is at least as large as the one given by the initial estimator (AGNES), and it is used to stabilize the trade-off between false positive and true positive edges in the original estimator (AGNES) when $n$ is small (for details see Section \ref{ROCc}). Moreover, as the sample size increases,  the value of $\lambda$ chosen by the A-MSE method tends to the original estimator of $\lambda$ (AGNES).  We use Path connectivity as the initial good choice of $\lambda$ to find the most sparse graph that is easy to interpret. Starting from the sparsest graph and proceeding to denser graph structures,  the PC method monitors the first big change in connectivity of the estimated networks, which is frequently associated with cluster agglomerations.

\begin{table}[h]
\scriptsize
\begin{center}
\caption{{\footnotesize Risk functions main characteristics. }} 
\begin{tabular}{ l| r r r r r r}
method & penalized & uses network  &subsampling & fully & fast& very sparse\\
& likelihood&characteristics.&& automatic&&graph estimates\\
\hline
PC &           &\checkmark &           &        \checkmark  &\checkmark &\checkmark  \\
A-MSE&        &\checkmark &\checkmark &\checkmark&          &\checkmark  \\
AGNES&         &\checkmark &           &\checkmark&\checkmark&  \\
StARS &           &\checkmark &\checkmark &          &          &  \\
BIC &\checkmark&           &           &\checkmark&  \checkmark        &\checkmark  \\
AIC &\checkmark&           &           &\checkmark& \checkmark         &  \\
\end{tabular}
\label{tabPOWER9}
\end{center}
\end{table}


\vspace{-0.7cm}
\section{Algorithms}\label{SEC3}

\subsection{Path connectivity regularization parameter selection}
The procedure to select $\lambda$ by Path connectivity is detailed in Algorithm \ref{algPC}. It is generally fast and straightforward, i.e. does not require any additional tuning.

\begin{algorithm}
\caption{Path connectivity algorithm}\label{algPC}
\begin{algorithmic}[1]
 \Procedure{R$_{PC}$}{$\lambda$}
\BState Set $\Lambda = (\lambda_k)_{k=1}^M$ with $\lambda_k - \lambda_{k-1} = h$, $k=2, \ldots, M$.
\For {k in 1 until M}:
\State Estimate the graph $\hat{G}^{\lambda_{k}}$ using \eqref{eq:ML2} and calculate its geodesic distance matrix 

\hspace{0.5cm}$[\hat{\dg}_{ij}]$ as in \eqref{eq:AG1}.
\State Calculate geodesic distance mean  $H(\lambda_{k})= m^{-1} \sum_{i<j} \hat{\dg}_{ij}(\lambda_{k})  I(\hat{\dg}_{ij}(\lambda_{k}) < \infty)$ 

\hspace{0.5cm}with $m = p(p-1)/2$.
\EndFor
\BState Calculate $D_h(\lambda_k) =  H(\lambda_k) - H(\lambda_{k-1})$ and  the running average $\bar{D}_h(\lambda_k) = 1/(M-k-1) \sum_{j=k}^M D_h(\lambda_j)$ for $(\lambda_k)_{k=2}^M$.
\BState Return $D_h(\lambda_k)/\bar{D}_h(\lambda_k)$, $k=2,\ldots,M$.

 \EndProcedure
\end{algorithmic}
\end{algorithm}
\vspace{-0.5cm}
\subsection{A-MSE regularization parameter selection}\label{AAG}
For $q=2$, the risk function $R_{AMSE}(\lambda)$ presented in (\ref{eq:AAGq}) can be decomposed by the sum of the variance and the squared bias, with  the corresponding approximation given by
\begin{equation}\label{eq:AAG2}
\hat{R}_{AMSE}(\lambda) = \sum_{i>j}  [\hat{\mathbf{E}}(\hat{\mathbf{E}}[\hat{d}_{ij}(\lambda)]-\hat{d}_{ij}(\lambda))^2 +
(\hat{\mathbf{E}}[\hat{d}_{ij}(\lambda)]-\hat{d}_{ij}(\lambda_{ac}))^2].
\end{equation}
Here $\hat{\mathbf{E}}(\hat{\mathbf{E}}[\hat{d}_{ij}(\lambda)]-\hat{d}_{ij}(\lambda))^2$ and $\hat{\mathbf{E}}[\hat{d}_{ij}(\lambda)]-\hat{d}_{ij}(\lambda_{ac})$ are estimators of the variance of $\hat{d}_{ij}(\lambda)$ and the bias of $\hat{d}_{ij}(\lambda)$ with respect to  $\hat{d}_{ij}(\lambda_{ac})$ using subsampling. The subsampling procedure to select $\lambda_{amse}$ is presented in Algorithm \ref{subsAGNES}. Following \cite{Meinshausen2010} we choose the effective sample size $B=0.5n$ since the procedure gets the closest to bootstrap. Nevertheless, other effective sizes could be used. For instance, \cite{Liu2011} use $B=10\sqrt{n}$.

\begin{algorithm}
\caption{Subsampling approach to approximate (\ref{eq:AAG2})}\label{subsAGNES}
\begin{algorithmic}[1]
 \Procedure{R$_{AMSE}$}{$\lambda$}
\BState Set $\Lambda = (\lambda_k)_{k=1}^M$ and number of subsampling replicates T.
\For {t in 1 until T}:
\State Subsample $B \subset \{1:n\}$ and set $X_B = (X_j, j \in B)$.
\State Estimate the graphs $\hat{G}^t(\lambda_k)$ for all $\lambda_k\in \Lambda$ using $X_B$.
\State Find dissimilarities of $\hat{G}^t(\lambda_k)$ by $
\hat{d}_{ij}^t(\lambda_k) = 1-\eta_{ij}^t(\lambda_k)/\sqrt{\kappa_i^t(\lambda_k)\kappa_j^t(\lambda_k)}$.
\EndFor
\BState Estimate the average $\bar{d}_{ij}(\lambda_k)$ over all $T$  iterations.
\BState Return  $T^{-1} \sum_{t=1}^T (\bar{d}_{ij}(\lambda_k)-\hat{d}_{ij}^t(\lambda_k))^2$ for all $\lambda_k\in \Lambda$.
 \EndProcedure
\end{algorithmic}
\end{algorithm}

\subsection{AGNES regularization parameter selection}

Below is the AGNES iterative clustering algorithm, including the agglomeration coefficient that is used to select $\lambda$. The input to the algorithm is a dissimilarity matrix $D = [d_{ij}] = \hat{D}(\lambda)$ based on the graph $\hat{G}^\lambda$ corresponding to the estimator $\hat\Omega^\lambda$ defined by \eqref{eq:ML2}.
AGNES performs hierarchical clustering by iteratively joining groups of nodes with the smallest average dissimilarity coefficient, starting with individual nodes as single clusters and finishing with a single cluster of all $p$ variables. Let $(C^{(t)}_1,\ldots,C^{(t)}_p)$ be a partition of $(1:p)$ at iteration $t$, and let $\delta_{k,\ell}^{(t)}$ denote  a dissimilarity between clusters $C^{(t)}_k$ and $C^{(t)}_m$.  We also record the dissimilarity for each node when it merges with another cluster or node for the first time, denoting it by $\delta^\star_j$, $j=1,\ldots,p$,  and the distance $\delta^\star_{\max}$ between the two clusters merged at the last step into the single cluster. The procedure is detailed in Algorithm \ref{AGNESalg}.

\begin{algorithm}[h]
\caption{AGNES clustering algorithm}\label{AGNESalg}
\begin{algorithmic}[1]
 \Procedure{R$_{AGNES}$}{$\lambda$}
\BState Initialization: take each node as an individual cluster, i.e. set $C_k^{(0)}=\{k\}$, $k=1,\ldots,p$, and $\delta^{(0)}_{k,\ell}= d_{k,\ell}$ - dissimilarity between nodes $k$ and $\ell$. 
     
\BState At iteration $t\geq 0$:
\State Find pair of clusters $(h,k)$ ($h<k$) with the smallest dissimilarity, i.e.
$$
(h,k) = \arg \min_{i<j}  \delta_{i,j}^{(t)}, 
$$

\hspace{-0.2cm}merge them, i.e. set $C_{k}^{(t+1)} = \{C_{k}^{(t)}, C_{h}^{(t)}\}$ and  remove cluster $h$: $C_{h}^{(t+1)}=\emptyset$. 

\hspace{-0.2cm}Remaining clusters are unchanged: 
set $C_{j}^{(t+1)}=C_{j}^{(t)}$ for $j\neq k,h$.


\State The dissimilarities change to 
$$
\delta^{(t+1)}_{j,h}=\delta^{(t+1)}_{h,j}=\infty, \quad 
\delta^{(t+1)}_{k,j}=\delta^{(t+1)}_{j,k}= \frac 1 2 \left[\delta^{(t)}_{k,j} + \delta^{(t)}_{j,h}\right], \quad \forall j\neq k,h.
$$

\hspace{-0.2cm}If $|C_{k}^{(t)}|=1$, set $\delta^\star_k=\delta_{k,h}^{(t)}$; if $|C_{h}^{(t)}|=1$, set $\delta^\star_h=\delta_{k,h}^{(t)}$.

\State If the number of non-empty sets (clusters) in the newly formed partition $(C^{(t+1)}_j)$ 

\hspace{-0.2cm}is more than 1, then set $t=t+1$ and go to step 3; otherwise set $\delta^\star_{\max}= \delta_{k,h}^{(t)}$ .

\BState Return
\begin{equation}\label{eq:AG2}
AC(\lambda) = \frac{1}{p} \sum_{j=1}^p \left(1- \frac{\delta^\star_j}{\delta^\star_{\max}}\right).
\end{equation}
 \EndProcedure
\end{algorithmic}
\end{algorithm}

The coefficient $AC(\lambda)$ measures the average distance between a node in the graph and its closest cluster of nodes. When the dissimilarities within the clusters are small in comparison to the maximum dissimilarity, then $1-\delta^\star_j/\delta^\star_{\max}$ is large for all $j$ and $AC(\lambda)$ is consequently high.

The time and total memory used in the AGNES algorithm increases exponentially as $p$ grows. In order to make computations feasible in very high dimensions, we use an approximation of the measure by a variable subset selection approach \citep{Kohavi1997}. We consider the average AC coefficient with respect to $\lambda$ over several sets of variables. We validate the subsets $V \subset \{1:p\}$ of size $|V|$ using the coefficients of variation of the empirical degree distribution ($\kappa$) defined by $\text{CV}_{V}= \text{sd}_{V}(\kappa)/\mathbf{E}_V(\kappa)$ with $\mathbf{E}_V(\kappa)=1/|V|\sum_{j\in V} \kappa_j $ and  $ \text{sd}_V(\kappa)= 1/(|V|-1)\sum_{j\in V} (\kappa_j-\mathbf{E}_V(\kappa))^2$ (see Algorithm \ref{subsetsAlg}). We aim to find subset of variables whose number of edges is approximately proportional to those in the original matrix.  In Section 4 of the supplementary material we illustrate how the variable subset approach reduces the computational time in high-dimensional simulated datasets.

\begin{algorithm}
\caption{Subset selection for AGNES computations}\label{subsetsAlg}
\begin{algorithmic}[1]
 \Procedure{S}{$\lambda$}
\BState Input: variables $V_t=\{1:p\}$ and their degrees $\kappa=\{\kappa_1,\ldots,\kappa_p\}$.
\BState Compute $\text{CV}_{V_t}$.
\BState Select randomly $m<p$ variables from the original data to form set $V_0\subset V_t$.
\BState Add all the nodes $V_1$ in the adjacency matrix $\hat{A}^{\lambda}$ which have a path to at least one node in $V_0$. Use $V_s=\{V_0,V_1\}$.
\BState  Compute $\text{CV}_{V_s}$. If $|\text{CV}_{V_s}/\text{CV}_{V_t} -1|>\tau$ go to step 4, otherwise return $V_s$. 
 \EndProcedure
\end{algorithmic}
\end{algorithm}
%


\section{Simulated data analysis}\label{SEC4}
In this section we consider simulated data to test the performance of the regularization parameter selection methods using graph structures similar to what can be expected in biological networks. We analyze both the capacity to obtain the true connections and the accuracy in recovering network characteristics of the true graph. 

\subsection{Graph topologies in biological data}\label{SEC22}
In real applications, the graph which defines causal connections between variables (e.g. genes, proteins, etc) is unknown but there is typically some knowledge about what kind of network structure can be expected \citep{Newman}. For instance, biological graph structures usually present associations in the shape of clusters, meaning that the nodes form groups that are more similar to the nodes within the group than to the nodes of other groups \citep{Eisen1998}. In addition, network patterns can be defined by the distribution of the variable $p_k$, which denotes the fraction of nodes in the network that has degree $k$. Here we consider two different graph topologies: hubs-based and power-law.

Hubs-based networks are graphs where only few nodes have a much higher degree (or connectivity) than the rest. This is a typical case in biological networks where nodes that behave as hubs may have different biological functions than the other nodes \citep{Lu2007}. Power-law networks assume that the variable $p_k$  follows a power-law distribution
$$
p_k =  k^{-\alpha}/\varsigma(\alpha),
$$
where $k \geq 1$, $\alpha$ is a positive constant and the normalizing function $\varsigma(\alpha)$ is the Riemann zeta function. Following \cite{Peng2009}, $\alpha = 2.3$ provides a distribution that is close to what is expected in biological networks.

\subsection{Simulated data }\label{simGM}
We generate data from multivariate normal distributions with zero mean vector and several almost-block diagonal precision matrices, where each block (or cluster) has a hubs-based or power-law underlying graph structure (defined in Section \ref{SEC22}) and there are some extra random connections between blocks. The non-zero partial correlation coefficients are simulated by
\begin{equation}
\Omega^{(0)}= [\omega_{ij}^{(0)}], \mbox{\hspace{0.5cm}} \omega_{ij}^{(0)} = \left\{ \begin{array}{r l l}
      & \text{Unif}(0.5,0.9) & \mbox{if $E_{ij} = 1$ with prob$=0.5$ } ;\\
      &  \text{Unif}(-0.5,-0.9) & \mbox{if $E_{ij} = 1$ with prob$=0.5$ }; \\
      &   0 & \mbox{if $E_{ij} = 0$}.
         \end{array} \right.
\label{eqSM11}
\end{equation}
Then, we regularize $\Omega^{(0)}$, which may not be positive definite, by  $\Omega^{(1)} = \Omega^{(0)} +  \delta I$, with $\delta$ such that the condition number of  $\Omega^{(1)}$ is less than the number of nodes, so obtaining a positive definite matrix \citep{Ai2011}.  Note that such precision matrices are non-singular, sparse and with the non-zero elements bounded away from 0.

We consider precision matrices with $p = $ 50, 170, 290 and 500 and sample sizes $n = $ 50, 100, 200, 500. Different number of hubs, degree of hubs, and sparsity levels are considered in 60 simulated datasets for each combination of $p$ and $n$. Full specification of simulated data is given in the supplementary material.

We use the R package \texttt{huge} \citep{Zhao2012a}  to estimate CD graph structures by GLasso and Neighborhood selection (MB). The GLasso gives the estimated partial correlation matrix but MB only provides the estimated adjacency matrix. In order to compare the proposed methods to both AIC and BIC, here we only present the results for the GLasso procedure. Nevertheless, the performance of the methods using MB estimates is shown in the supplementary material. We take a sequence of 70 equidistant points for $\lambda$ going from $0.20$ to $0.66$ for small $n$ and a sequence going from $0.03$ to $0.40$ for large $n$ (the graphs almost have no change for $\lambda$'s smaller than the lower limit with all nodes connected as well as higher than the upper limit with no edges across nodes).  Then we select $\lambda$ by six different approaches:  1) PC; 2) A-MSE; 3) AGNES; 4) StARS; 5) BIC and 6) AIC. StARS (with $\beta=0.05$) produces the lowest $\lambda$ for almost all the simulated datasets followed closely by AIC. The BIC results are strongly dependent on the sample size; the methods selects large tuning parameters for small $n$ and low tuning parameters for large $n$ in comparison to A-MSE. The AGNES selections are always larger than A-MSE but they get close when $n$ increases. The PC $\lambda$ selections do not vary much for different $n$ and $p$ scenarios and produce similar magnitudes to $\lambda$'s selected by A-MSE.

We assess the performance of the $\lambda$ selection approaches for GLasso estimates using two different measures: squared errors in both the partial correlation matrix and the dissimilarity matrix defined in (\ref{eq:AG1}) and graph recovery with a false positive and true positive analysis. The simulated data analysis is completed in the supplementary material where we compare for both GLasso and MB the selected graph structures and the true networks given global network characteristics as clustering, connectivity and graph topology.

\subsection{Mean square errors}\label{MSE}
To measure performance of the methods we use the ranks of the average mean square errors (MSE) of the partial correlation matrix $\Omega$ (Table \ref{tabPOWER3}) as well as of the dissimilarity matrix $D$ (Table \ref{tabPOWER1}).
This second rate gives a good reference to determine if the estimated graph captures  the true local structure.
The lowest rank (rank = 1) is assigned to the lowest MSE and the largest rank (rank = 6) is for the largest MSE out of the six approaches. In the tables, we show the errors for the GLasso method. 

Even though StARS and AIC estimate $\Omega$ well, they produce larger errors than AGNES, A-MSE, PC and BIC when minimizing the MSE of the dissimilarity matrix. Particularly, A-MSE tends to be the best selection for this loss function for large $n$. We find that BIC does well for small $n$, contrarily of what is obtained in \cite{Liu2011}, but tends to be unreliable for larger sample sizes. AGNES gives good ranks for the hubs-based scenarios, particularly when $n$ is large, and PC is almost always among the three best methods.

\begin{table}[h]
\scriptsize
\begin{center}
\caption{{\footnotesize Average ranks for the mean square errors of the precision matrix.} }  
\begin{tabular}{ l r r r r  r  r r r r}
&\multicolumn{4}{c}{Hubs-based}&&\multicolumn{4}{c}{Power law}\\
\cline{2-5}\cline{7-10}
n & $50$ &$100$& $200$ & $500$ & &$50$ &$100$& $200$ & $500$\\
\hline
\multicolumn{10}{c}{dimension p=50} \\
AGNES  &3.05 &3.55 &4.06 &4.40 &&3.12 &3.73 &4.40 &4.71\\
A-MSE&4.33 &4.90 &5.22 &5.38 &&4.92 &5.47 &5.67 &5.78\\
PC     &5.23 &5.80 &5.58 &5.15 &&4.58 &5.13 &4.85 &4.49\\
StARS  &\textbf{1.27} &\textbf{1.49}& \textbf{1.18}& \textbf{1.28} &&\textbf{1.17} &\textbf{1.43} &\textbf{1.04} &\textbf{1.07}\\
BIC    &5.38 &3.73& 3.14& 3.06 &&5.33 &3.66 &3.08 &3.02\\
AIC    &1.73 &1.52& 1.82& 1.73 &&1.90 &1.58 &1.96 &1.92\\
\multicolumn{10}{c}{dimension p=170} \\
AGNES  &2.81 &3.30 &4.07 &4.63 &&2.69 &3.28 &3.91 &4.42\\
A-MSE&4.13 &4.92 &5.17 &5.52 &&4.79 &5.38 &5.43 &5.63\\
PC     &5.31 &5.91 &5.61 &4.03 &&4.96 &5.55 &5.46 &4.93\\
StARS  &\textbf{1.00} &\textbf{1.22} &\textbf{1.00} &\textbf{1.00} &&\textbf{1.00} &\textbf{1.00} &\textbf{1.00} &\textbf{1.05}\\
BIC    &5.56 &3.87 &3.14 &3.48 &&5.23 &3.78 &3.20 &3.00\\
AIC    &2.19 &1.78 &2.02 &2.35 &&2.33 &2.00 &2.00 &1.97\\
\multicolumn{10}{c}{dimension p=290} \\
AGNES  &2.54& 3.02& 3.97& 4.38&&2.33& 3.07& 3.87& 4.20\\
A-MSE&4.17& 4.83& 5.08& 5.30&&4.83& 5.39& 5.46& 5.46\\
PC     &5.12& 5.91& 5.78& 4.83&&4.68& 5.57& 5.53& 5.33\\
StARS  &\textbf{1.00}& \textbf{1.01}& \textbf{1.00}& \textbf{1.00} &&\textbf{1.00}& \textbf{1.00}& \textbf{1.00}& \textbf{1.02}\\
BIC    &5.71& 4.23& 3.13& 3.29&&5.47& 3.98& 3.14& 3.02\\
AIC    &2.46& 1.99& 2.03& 2.20&&2.68& 2.00& 2.00& 1.98\\
\multicolumn{10}{c}{dimension p=500} \\
AGNES  &2.13 &3.00&3.92&4.30 &&2.11& 3.00& 3.62&4.11\\
A-MSE&4.28 &4.78&5.13&5.35 &&4.81& 5.25& 5.27&5.47\\
PC     &4.94 &5.97&5.60&4.85 &&4.63& 5.67& 5.73&5.39\\
StARS  &\textbf{1.00}& \textbf{1.01}& \textbf{1.00}& \textbf{1.00} &&\textbf{1.00}& \textbf{1.00}& \textbf{1.00}& \textbf{1.00}\\
BIC    &5.78 &4.25&3.31&3.32 &&5.55& 4.08& 3.38&3.03\\
AIC    &2.88 &2.00&2.05&2.18 &&2.90& 2.00& 2.00&2.00\\
\end{tabular}
\label{tabPOWER3}
\end{center}
\end{table}

\begin{table}[ht]
\scriptsize
\begin{center}
\caption{{\footnotesize Average ranks for the mean square errors of the dissimilarity matrix}. } 
\begin{tabular}{ l r r r r  r  r r r r}
&\multicolumn{4}{c}{Hubs-based}&&\multicolumn{4}{c}{Power law}\\
\cline{2-5}\cline{7-10}
n & $50$ &$100$& $200$ & $500$ & &$50$ &$100$& $200$ & $500$\\
\hline
\multicolumn{10}{c}{dimension p=50} \\
AGNES  & 2.88 &2.70 &2.23 &2.09 && 3.60& 3.02& 2.38& 2.09\\
A-MSE& \textbf{2.83} &\textbf{2.47} &\textbf{1.65} &\textbf{1.44} && \textbf{2.12}& \textbf{1.65}& \textbf{1.20}& \textbf{1.41}\\
PC     & 3.52 &3.67 &2.75 &2.68 && 2.42& 2.22& 2.53& 2.52\\
StARS  & 4.16 &4.58 &5.81 &5.72 && 5.70& 5.58& 5.97& 5.96\\
BIC    & 3.83 &3.05 &3.41 &3.79 && 2.13& 3.12& 3.88& 3.98\\
AIC    & 3.77 &4.54 &5.16 &5.28 && 5.02& 5.42& 5.03& 5.04\\
\multicolumn{10}{c}{dimension p=170} \\
AGNES  &3.52 &2.98 &2.12 &1.73 && 4.31& 3.68& 3.06& 2.32\\
A-MSE&2.62 &\textbf{2.04} &\textbf{1.65} &\textbf{1.45} && 2.14& \textbf{1.58}& \textbf{1.45}& \textbf{1.40}\\
PC     &2.46 &2.49 &3.11 &3.83 && 2.12& 1.62& 1.73& 2.32\\
StARS  &6.00 &5.78 &6.00 &6.00 && 6.00& 6.00& 6.00& 6.00\\
BIC    &\textbf{2.14} &2.52 &3.14 &3.34 && \textbf{1.80}& 3.12& 3.77& 3.97\\
AIC    &4.26 &5.18 &4.98 &4.65 && 4.62& 5.00& 5.00& 5.00\\
\multicolumn{10}{c}{dimension p=290} \\
AGNES  &4.16& 3.02& \textbf{1.83}& 1.77 &&4.67& 3.92& 3.08& 2.57\\
A-MSE&2.42& \textbf{1.97}& 1.85& \textbf{1.48} &&2.17& 1.57& \textbf{1.53}& \textbf{1.48}\\
PC     &2.11& 2.89& 3.50& 3.66 &&2.30& \textbf{1.50}& 1.54& 1.98\\
StARS  &6.00& 5.99& 6.00& 6.00 &&6.00& 6.00& 6.00& 6.00\\
BIC    &\textbf{2.01}& 2.22& 2.85& 3.29 &&\textbf{1.55}& 3.01& 3.84& 3.98\\
AIC    &4.31& 4.91& 4.97& 4.80 &&4.32& 5.00& 5.00& 5.00\\
\multicolumn{10}{c}{dimension p=500} \\
AGNES  &4.83 &3.25&2.06&\textbf{1.60} &&4.89& 4.00& 3.38&2.56\\
A-MSE&2.51 &\textbf{1.80}&\textbf{1.94}&2.00 &&2.09& 1.72& \textbf{1.33}&\textbf{1.42}\\
PC     &2.04 &3.12&3.69&3.83 &&2.32& \textbf{1.48}& 1.68&2.06\\
StARS  &6.00 &6.00&6.00&6.00 &&6.00& 6.00& 6.00&6.00\\
BIC    &\textbf{1.51} &1.95&2.40&2.78 &&\textbf{1.60}& 2.80& 3.61&3.97\\
AIC    &4.11 &4.88&4.92&4.79 &&4.10& 5.00& 5.00&5.00\\
\end{tabular}
\label{tabPOWER1}
\end{center}
\end{table}


\subsection{Graph recovery}\label{ROCc}
\begin{figure}[ht]
\begin{center}
\includegraphics[width=14cm,height=7cm]{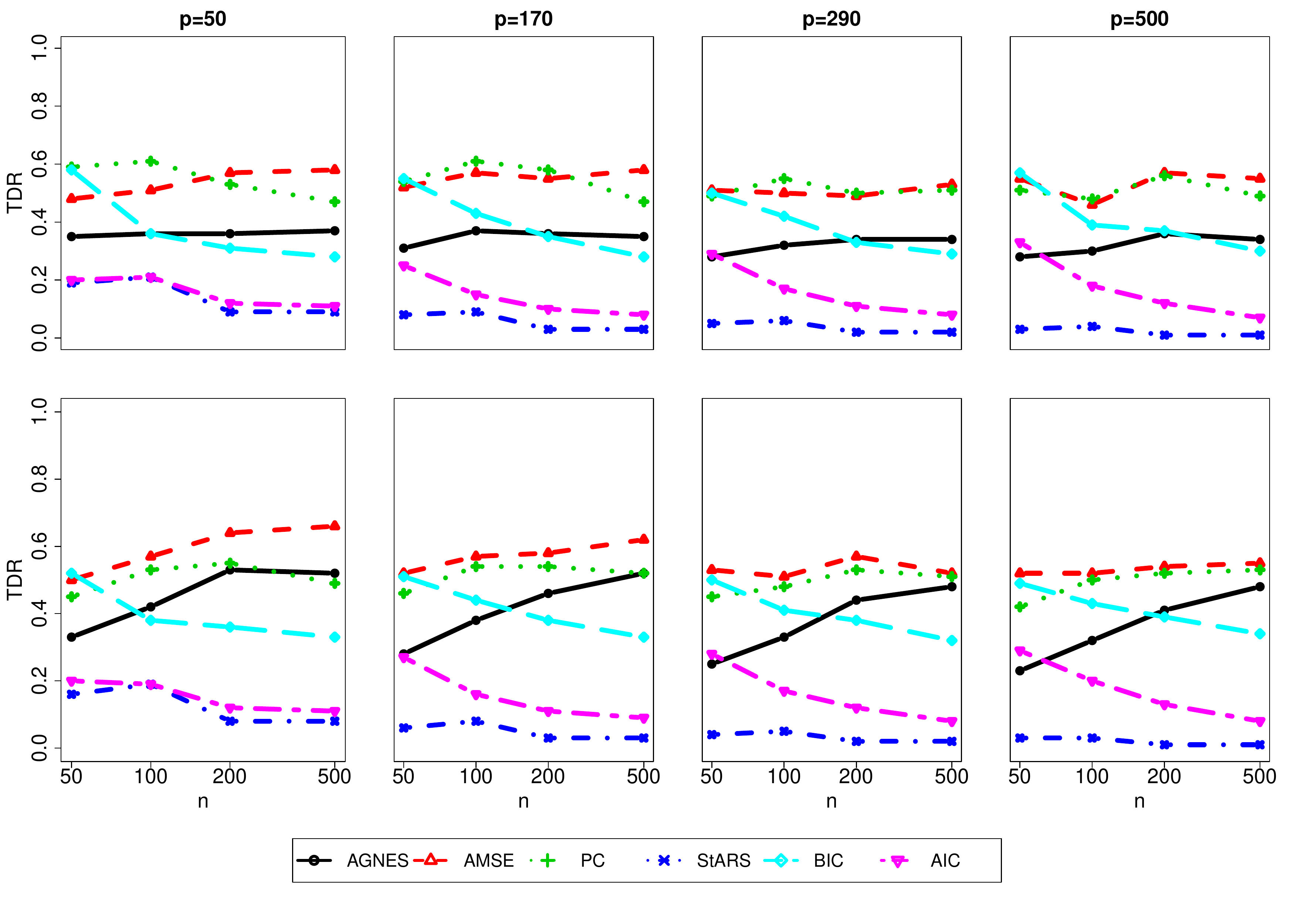}
 \caption{\footnotesize{True discovery rate for all $\lambda$ selection approaches and all combinations of $p$ and $n$. The x-axis scale is $n : \log(n)$}}\label{tdr}
\end{center}
\end{figure}

In order to quantify how well the algorithms recover the non-zero elements in $\Omega$ we compare the true discovery rate (TDR), which can be defined by $TDR =  TP/(TP+FP)$ with
$$
TP = \sum_{i<j}I(\hat{\Omega}_{ij} \neq 0 \mbox{\hspace{0.1cm} and \hspace{0.1cm} } \Omega_{ij} \neq 0), \mbox{\hspace{.5cm}} FP = \sum_{i<j}I(\hat{\Omega}_{ij} \neq 0 \mbox{\hspace{0.1cm} and \hspace{0.1cm} } \Omega_{ij} = 0), 
$$
for each of the estimated networks. In Figure \ref{tdr}, we show the average TDR in the 60 simulations for all considered combinations of $n$ and $p$. The TDR increases with $n$ for AGNES, A-MSE and PC whereas for AIC and BIC it goes down.  In this analysis we can see the limitations of the BIC method whose main goal is not the graph recovery of $\Omega$. BIC passes from selecting very sparse graphs with more TP than FP when $n$ is small to selecting much denser graphs with many more FP than TP when $n$ is large. 

\subsection{Summary}
In our simulations A-MSE turned out to be the best approach to recover the CD graph structure as can be seen in Table \ref{tabPOWER1}. BIC is also competitive when $n$ is small, but it is not reliable when analyzing larger sample sizes. PC is computationally the fastest method and only does slightly worse than A-MSE in Table \ref{tabPOWER1}.  Moreover, it generally obtains simple graph structures which result in comprehensible connectivity interpretations. The AGNES procedure is usually over-performed by the augmented version A-MSE for small $n$. For large $n$, AGNES and A-MSE have similar $\lambda$ selections with AGNES being significantly faster than A-MSE. AIC and StARS (using its default values) produce dense graph estimations and achieve the best results when minimizing the mean square error of $\Omega$. Nevertheless, they fail to obtain interpretable network structures due to poor graph recovery.


\section{Application to colon cancer gene expression data}\label{SEC5}
We apply the methods in a real case study. A gene expression data set that can be obtained from the TCGA platform at \url{https://tcga-data.nci.nih.gov/tcga/}. A total of 154 patients are examined, the gene expression profiling for 17,617 genes is obtained and normalized in each one of them for a colorectal tumor sample.

\begin{figure}[ht]
\begin{center}
 \begin{tabular}{cc}
    \subfloat[PC selected graph.]{\includegraphics[scale=0.22]{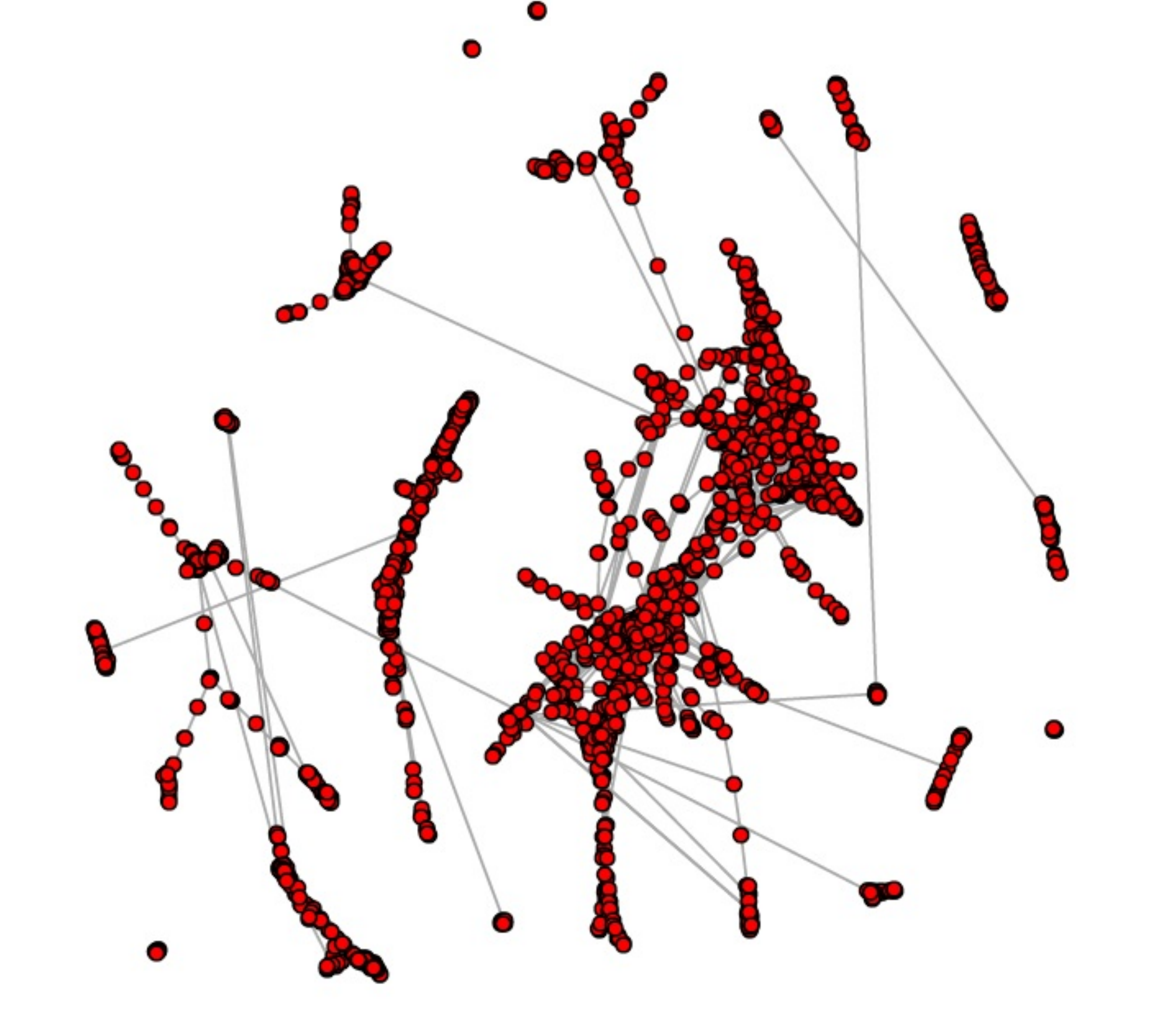}}
   & \subfloat[A-MSE selected graph.]{\includegraphics[scale=0.22]{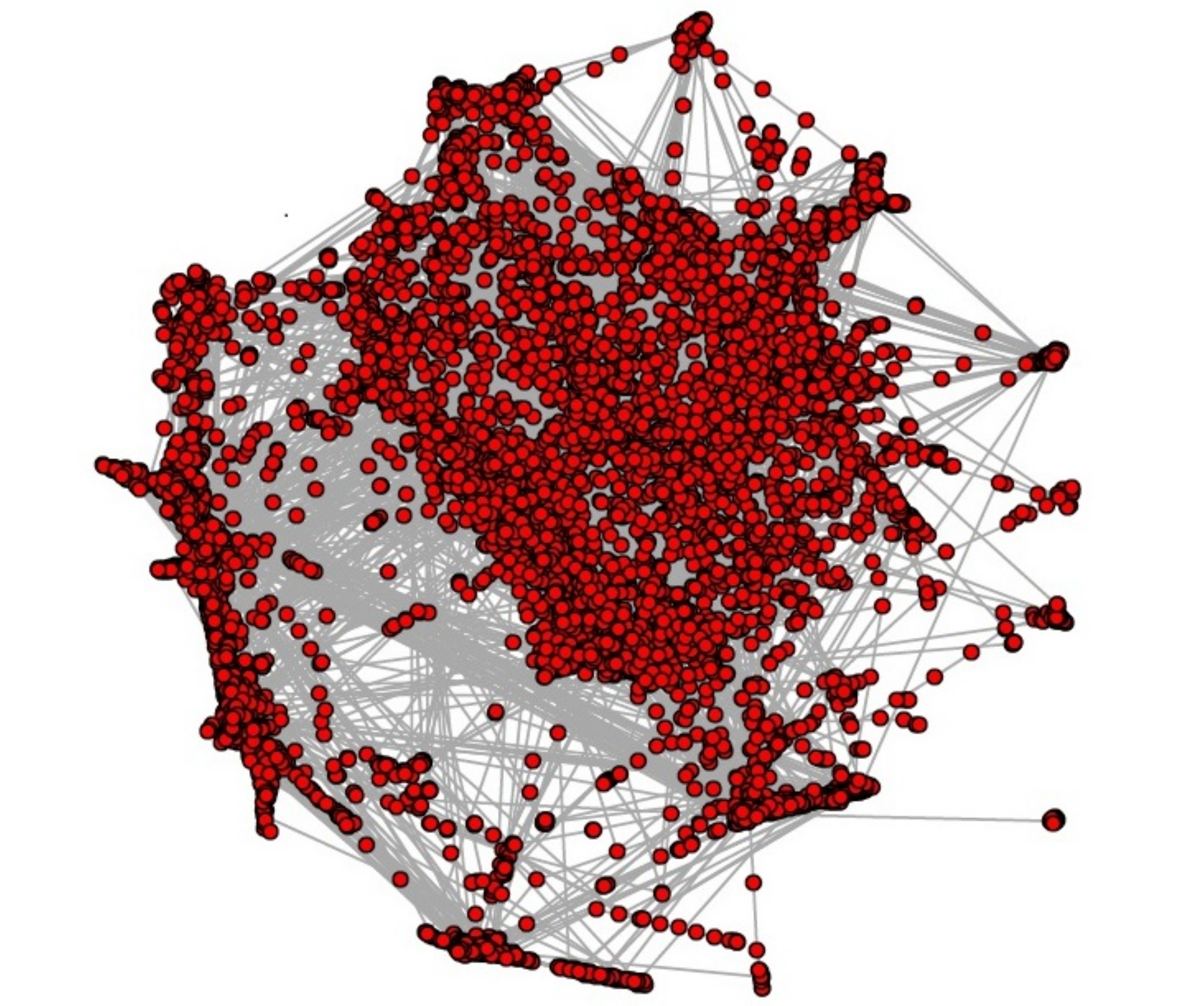}}\\
\end{tabular}
 \caption{\footnotesize{Best graphs by PC and A-MSE by using the estimated adjacency matrices by MB in the gene expression data.}}\label{freal}
\end{center}
\end{figure}

A reduction on the variable space is applied so that we only keep the most highly correlated genes. We use a filter for the gene's average square correlation with threshold equal to $0.04$. Moreover, we add the non-filtered genes which have at least one correlation coefficient with the filtered genes larger than $0.5$. This means a reduction to the $55\%$ of the genes with a total of 9,723 genes left to analyze. We estimate CD graphs via the Neighborhood selection algorithm of \cite{Meinshausen2006}.  We compute 90 different graphs given an equidistant sequence of $\lambda$'s between $0.35$ and $0.80$. Values of  $\lambda$ lower than $0.35$ produce almost-fully connected graphs and values above $0.80$ produce zero edges in the graph. We use the PC and A-MSE approaches to select one particular graph with $\lambda_{pc} = 0.69$ and $\lambda_{amse} = 0.55$. The graphical representation of the two underlying networks is presented in Figure \ref{freal}. The graph by PC, with $4,819$ edges, shows a simpler structure compared to A-MSE, with 19,986 edges.

We separate the graphs in different clusters by applying a Partitioning Around Medoids \citep{Reynolds2006} on the shortest distance matrix. We choose the number of clusters manually by considering the largest rate of change in the within-subject and between-subject variation such that the PC graph structure contains 15 clusters and the A-MSE contains 18 clusters.

To assess which biological processes may be linked to the clusters, we download 1,320  gene sets from the MSig database (http://www.broadinstitute.org/gsea/msigdb/index.jsp), which represent canonical pathways compiled from two sources: KeGG (\url{http://www.genome.jp/kegg/pathway.html}) and Reactome (\url{http://www.reactome.org/}). For each pathway we test for a significant over-representation in a cluster by using Fisher's exact test applied to the $2\times 2$-table defined by pathway and cluster membership with a Bonferroni correction for multiple testing. Note that we use the reduced selection of 9,723 genes here as ''background'', i.e.\ the analysis corrects for any over-representation of a pathway in that selection. 

In the PC selected graph, 6 out of 15 clusters have at least one significant pathway list (at 0.01 significant level) and a total of 160 significant pathway lists. Moreover, in the A-MSE graph, 7 out of 18 have also significant pathways and a total of 122 significant pathway lists. Among the significant lists, PLK1, NFAT, DNA replication or adaptive immune system are pathways associated with tumor cells.

\section{Discussion}\label{SEC6}
In this paper we study the problem of choosing the regularization parameter $\lambda$ for Gaussian graphical models in high dimensional data assuming we have high level knowledge about the nature of the graph structures,  namely strong clustering in the case of gene expression data \citep[e.g.][]{Eisen1998}.  The methods we introduce here take this assumption into account by selecting $\lambda$ so that risk functions measuring the degree of clustering (AGNES, A-MSE) or connectivity (PC) are optimized.
We aim to select the sparsest graph such that the real cluster structure is maintained and at the same time it contains a good tradeoff between true and false positive edges. The proposed approaches to select the regularization parameter provide competitive results at a relatively high computational speed. They present more reliable results than the StARS approach which tends to overestimate the network size. The StARS method accounts for the stability of the estimated graphs and has been proven to work well in \cite{Liu2011}. It depends, however, on another parameter which controls the maximum amount of variability in the graph. There is no straightforward choice for this parameter and our simulation study shows that using the default value of $0.05$ StARS yields uninformative networks with a majority of edges being false positives. 

The Path connectivity approach introduced here provides a good compromise between estimating the structure well and the number false positive edges. The main characteristic of this approach is that it relies on the shortest distance between all pairs of nodes. Interestingly, this quantity tends to show a clear changepoint when studied as a function of $\lambda$, at which the structure of the graph changes radically. 
%
It typically produces very informative graphs in all the tested simulated datasets and gives competitive results for the mean square error between dissimilarity matrices as discussed in Section \ref{MSE}. In the gene expression data set it also provides us with a clearly structured informative graph. PC gives an excellent first choice of $\lambda$ if we want to find an easily interpretable graph.

The A-MSE, with initial graph structure given by the AGNES selected graph, is the best of all the approaches in terms of minimizing the MSE between the true distances and the estimated ones in the simulated data. Also, $\lambda_{amse}$ is always smaller than $\lambda_{ac}$ leading to  less complex graphs  than the ones estimated by AGNES. This is a desirable property as we assume only a small proportion of non-zero elements in $\Omega$ and thus with increasing graph density the number of false positive edges grows much faster than the number of true positives. However, if the aim is to have fewer false negatives, that is, that as many as possible true edges are included at the expense of a higher number of false positives, then algorithms like AGNES and StARS are more appropriate.



The analysis of the gene expression data underlines some interesting results.
The obtained graphs present a cluster-based structure as we can see in Figure \ref{freal}.  Our new approach of choosing a regularization parameter, Path connectivity, leads to a sparse and clustered network that is easy to interpret.   Closer investigation of the results shows that the clusters overlap significantly with a number of pre-defined gene sets and regulatory pathways which indicates that our assumption of a sparse clustered structure leads us to biologically meaningful results.

In conclusion, we find that approaches such as PC, A-MSE and AGNES, which use network characteristics for parameter selection, can be beneficial in estimating CD graph structures (sparse partial correlation matrices) for high-dimensional biological data. While maintaining good statistical properties in terms of false discovery rates and mean square error, the resulting graphs tend to be easier to interpret from a biological perspective and thus are more useful in applications compared to parameter selection methods based on penalized log likelihood such as AIC or BIC.


\newpage
\section{Supplementary Material}
\subsection{Simulated data: graph information}\label{SUP:SEC1}
We generate data from a multivariate normal distribution using the next characteristics for the data:
\begin{enumerate}
\item 4 sample sizes: $n=50$, $n=100$, $n=200$, $n=500$.
\item 4 dimension sizes: $p=50$, $p=170$, $p=290$, $p=500$. 
\item Number of clusters (and variables per cluster) for each $p$ setting: 
      1 (50), 3 (70, 60, 40), 5 (70, 100, 40, 50, 30), 7 (100, 100, 80, 60, 60, 70, 30).
\item Hubs-based and Power-law models to generate data.
\item Probability for a variable to be a hub: $0.015$ (only for hubs-based models).
\item Degree of hub nodes generated by  Uniform$(5,b)$ where $b$ is one third of the number of variables per cluster.
\item Probability for presence of all remaining edges in hubs-based models: generated by Uniform$(0.005,0.03)$.
\item Probability for presence of edges in between clusters: generated by an Uniform$(0,0.1)$.
\item Parameter $\alpha$ in power-law models: $2.3$.
\end{enumerate}

\newpage
\onehalfspacing
\subsection{Simulated data: some of the generated graphs}
\begin{figure}[H]
\vspace{-1cm}
\begin{center}
 \begin{tabular}{cc}
\subfloat[p=50, hubs-based]{\includegraphics[height=3cm,width=5cm]{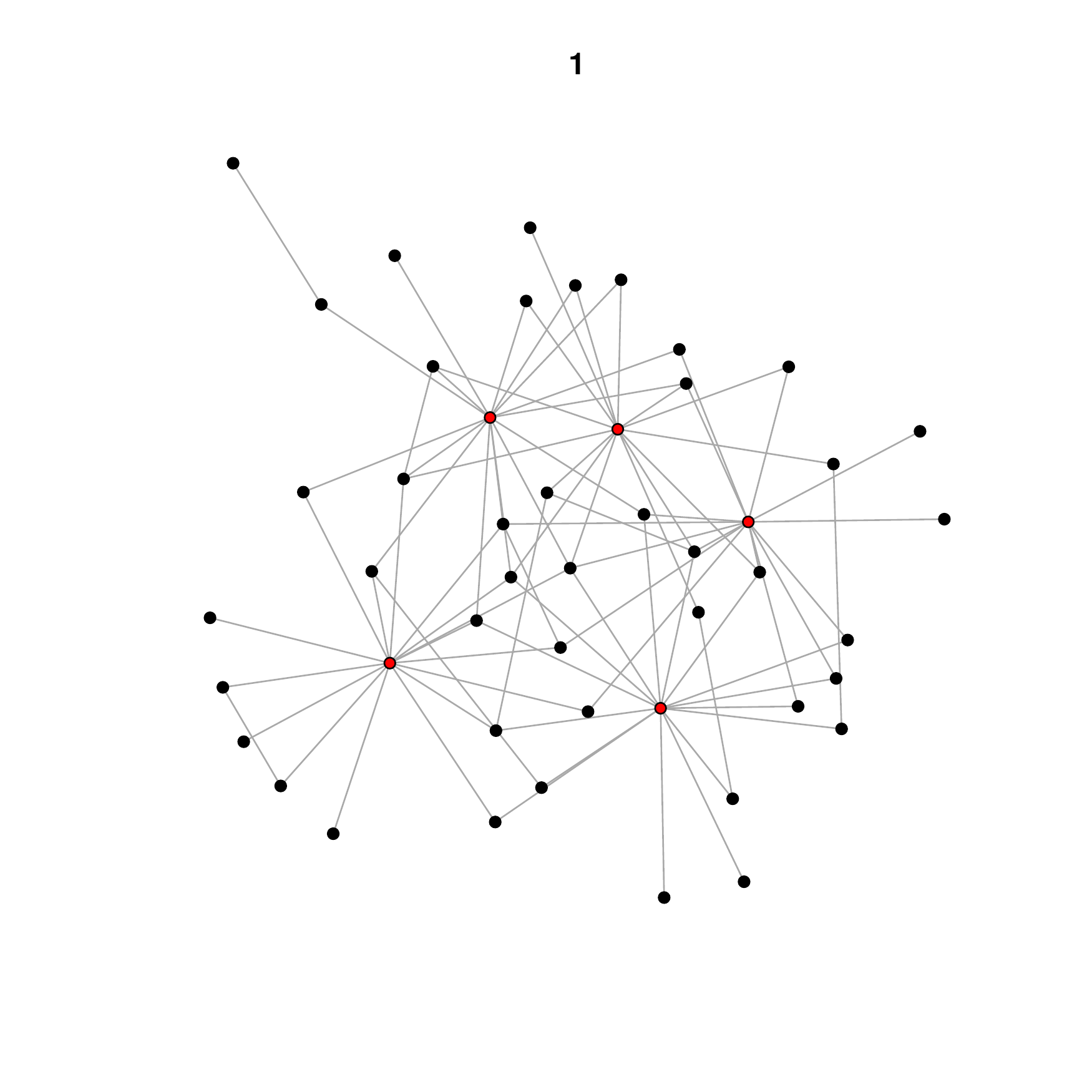}}&
\subfloat[p=50, power-law]{\includegraphics[height=3cm,width=5cm]{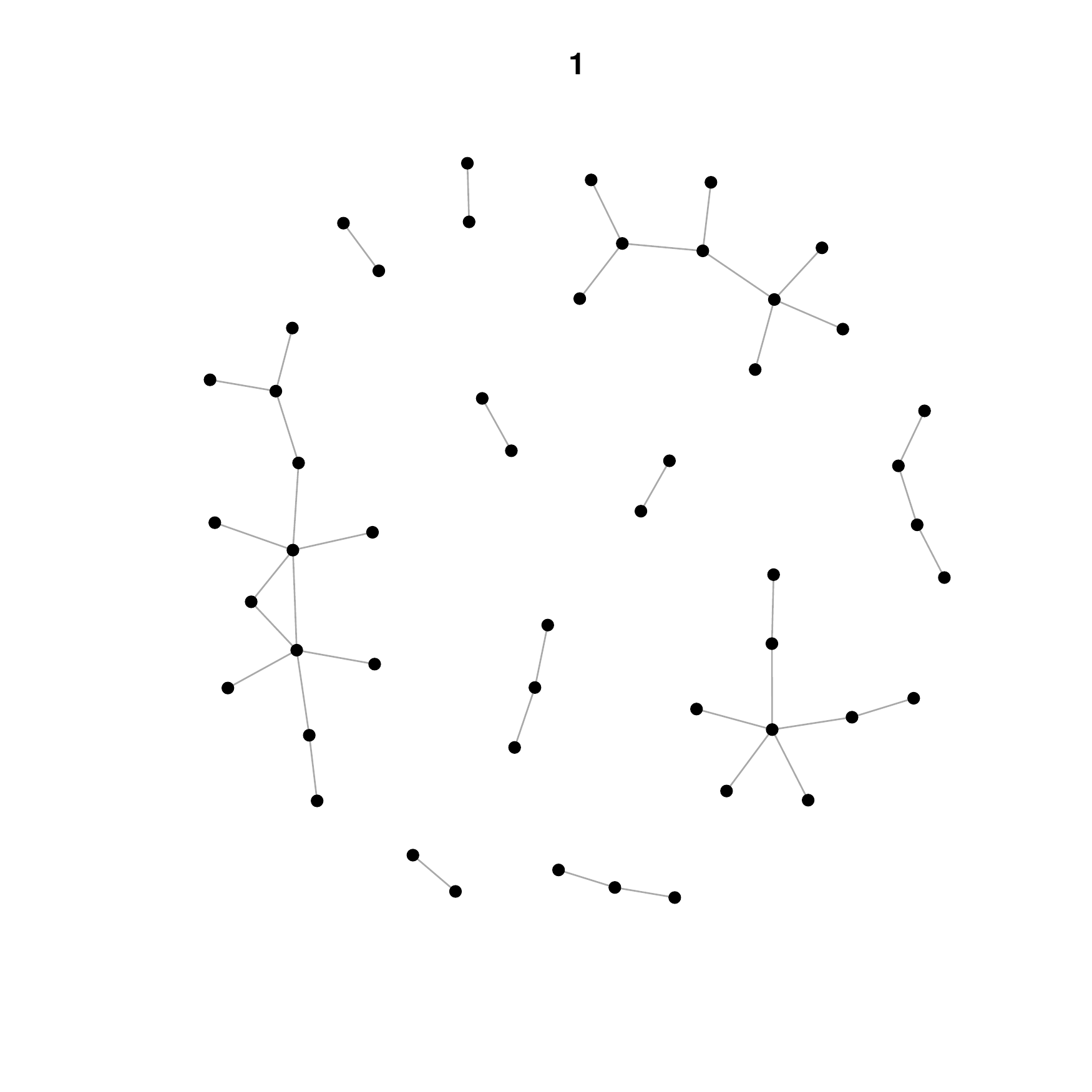}}\\
\subfloat[p=170, hubs-based]{\includegraphics[height=3cm,width=5cm]{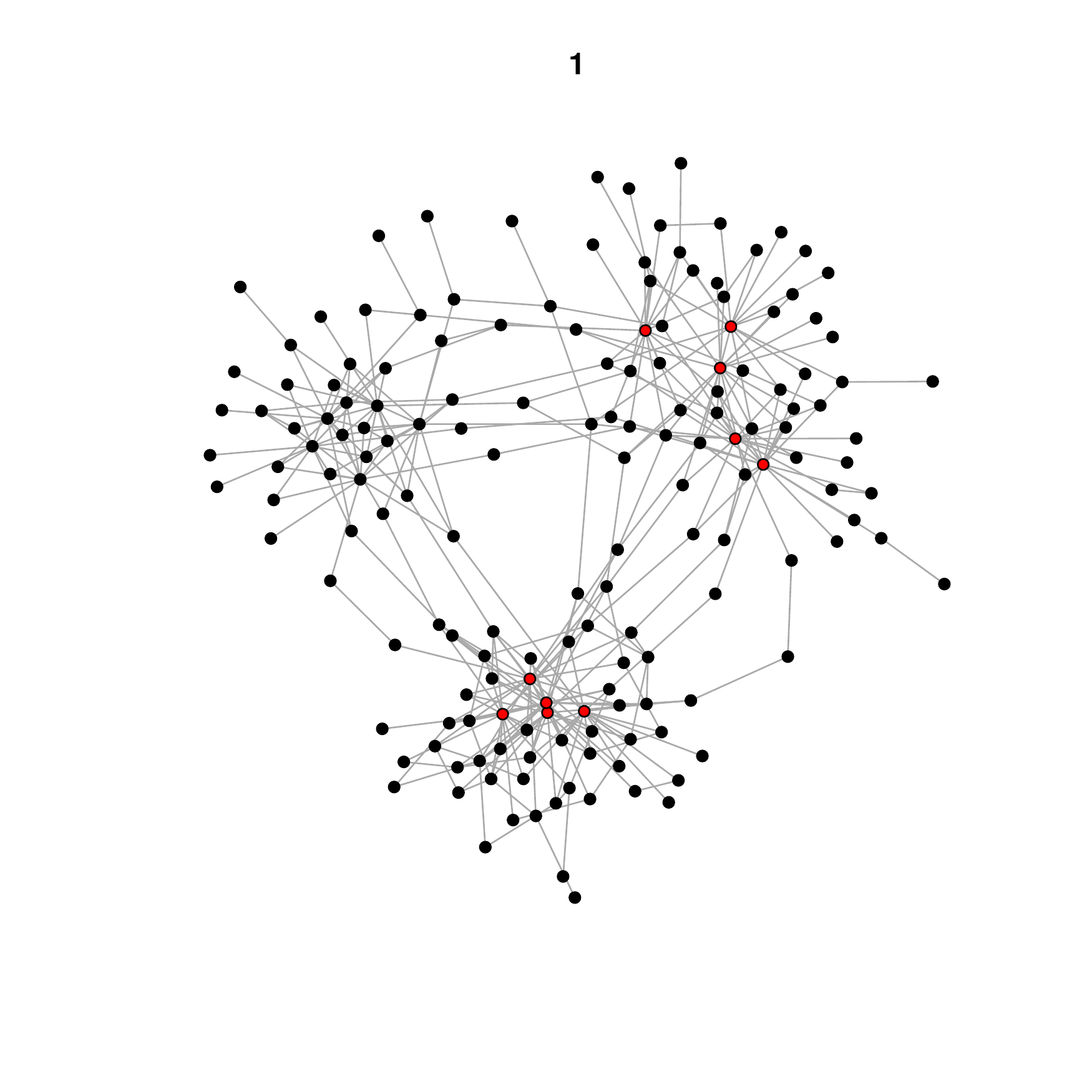}}&
\subfloat[p=170, power-law]{\includegraphics[height=3cm,width=5cm]{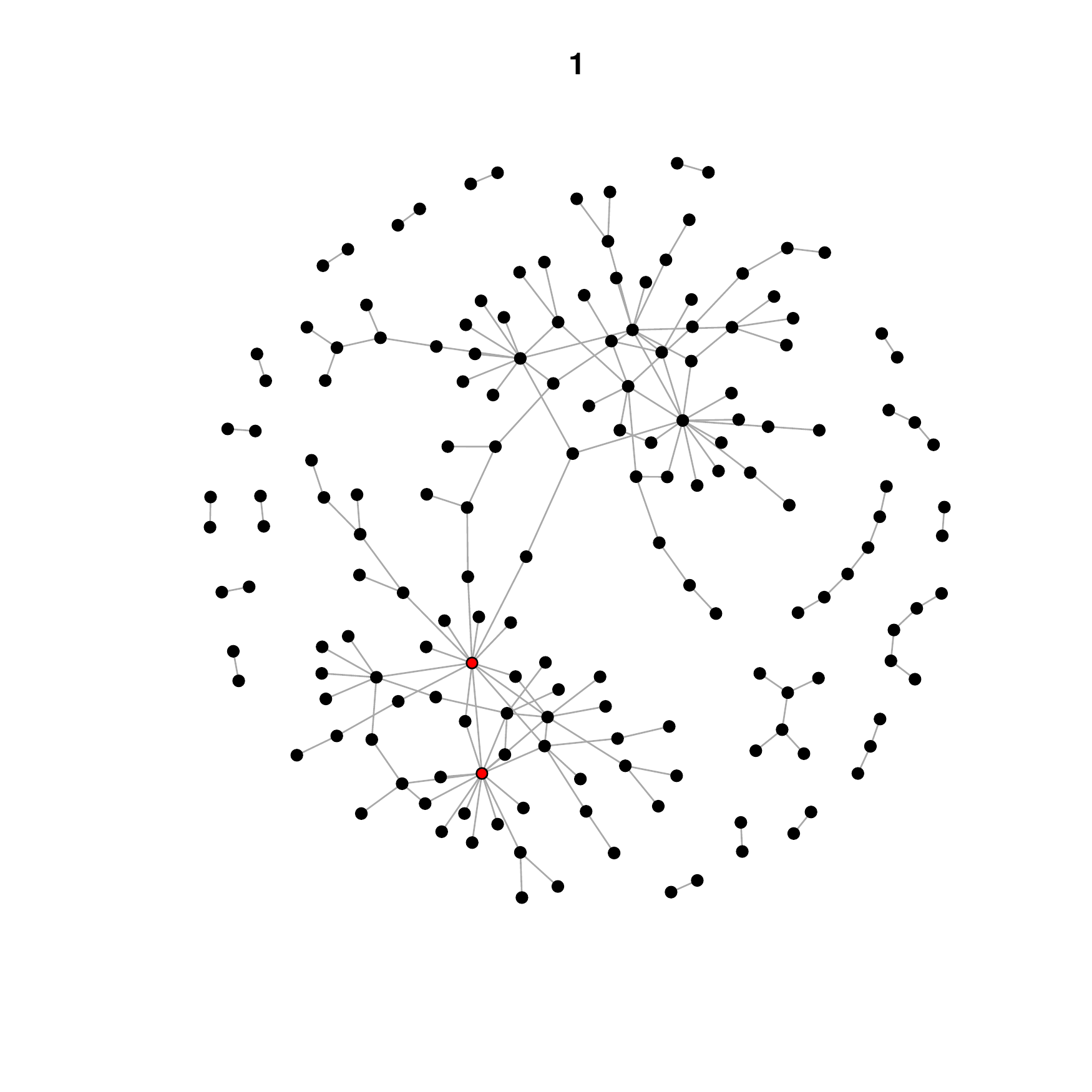}}\\
\subfloat[p=290, hubs-based]{\includegraphics[height=3cm,width=5cm]{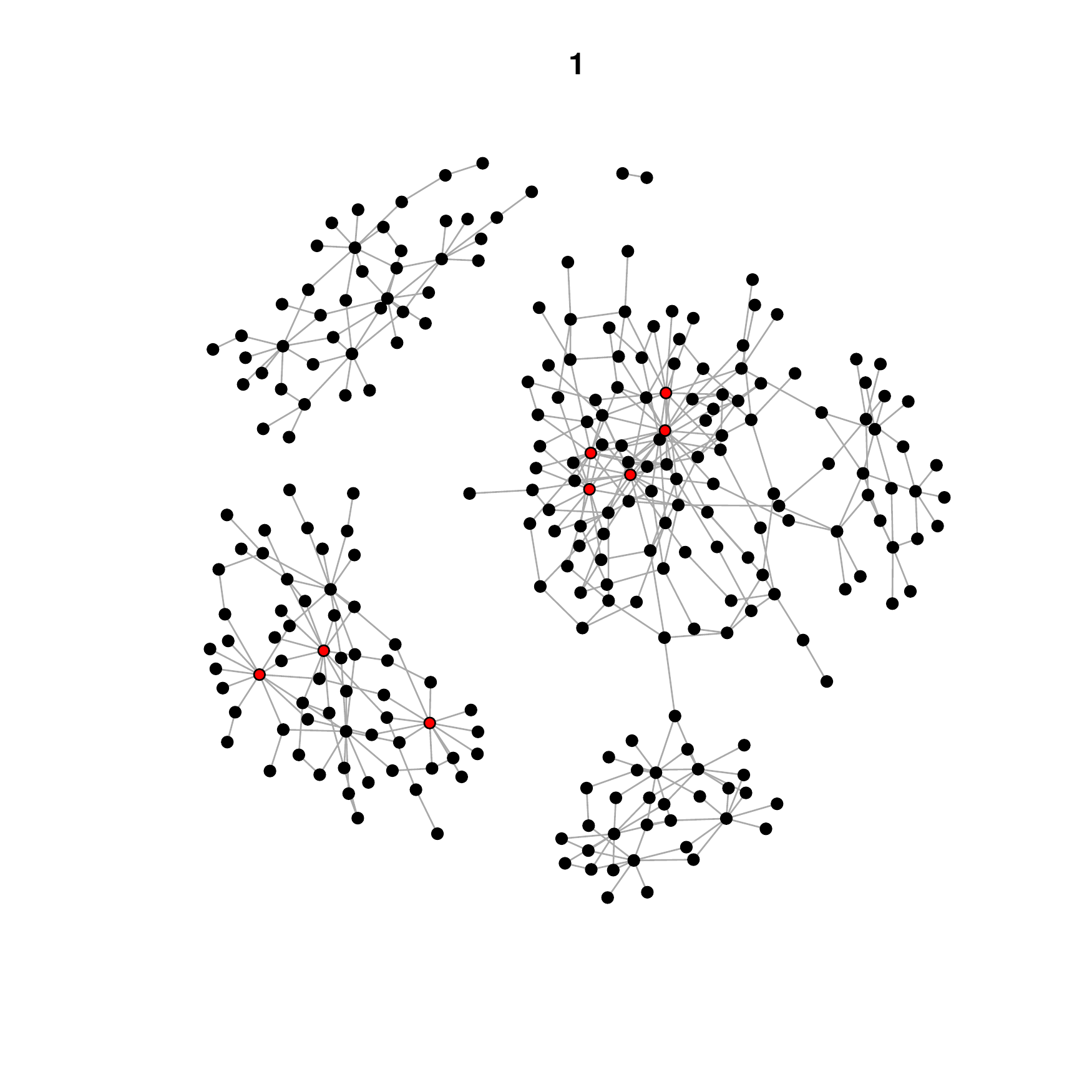}}&
\subfloat[p=290, power-law]{\includegraphics[height=3cm,width=5cm]{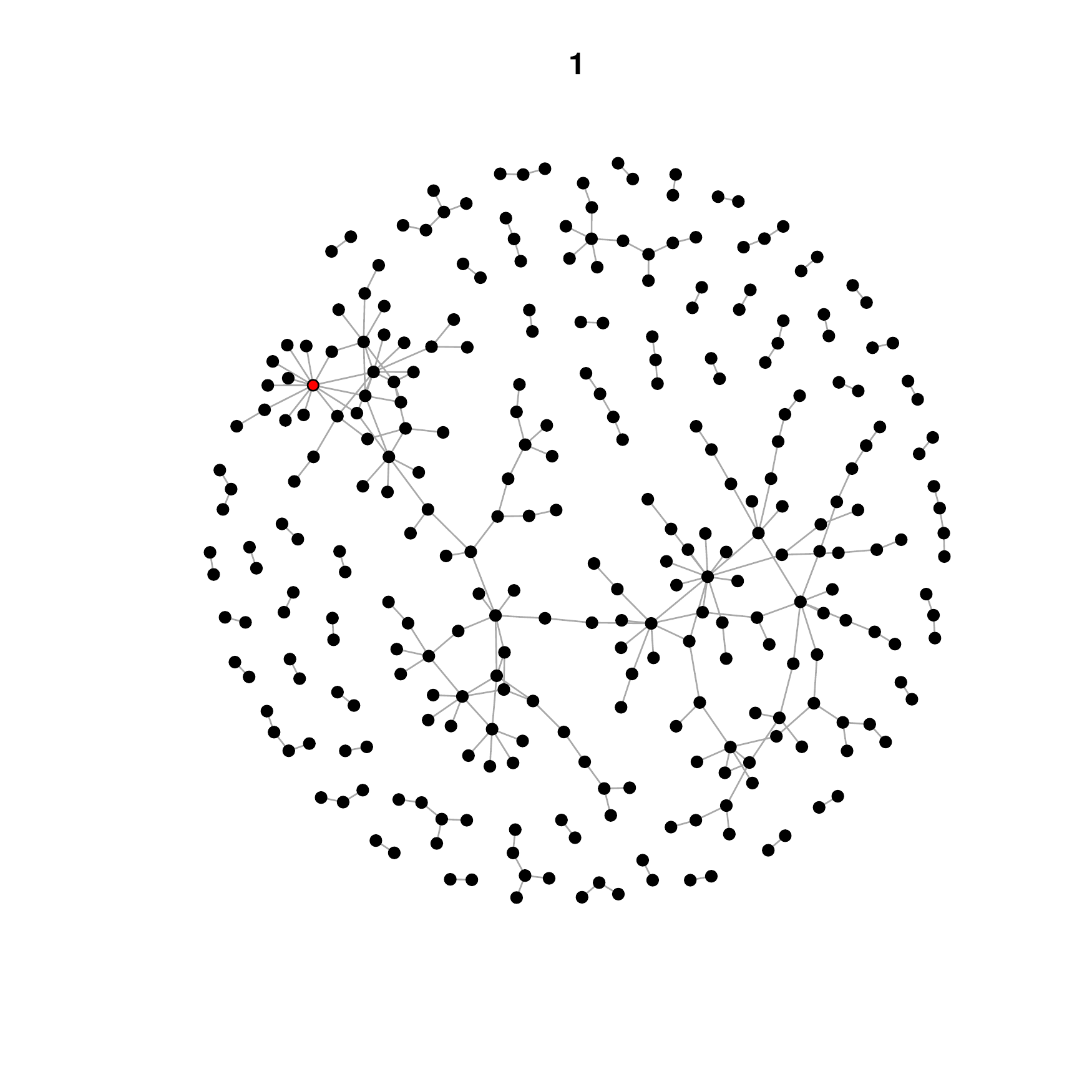}}\\
\subfloat[p=500, hubs-based]{\includegraphics[height=3cm,width=5cm]{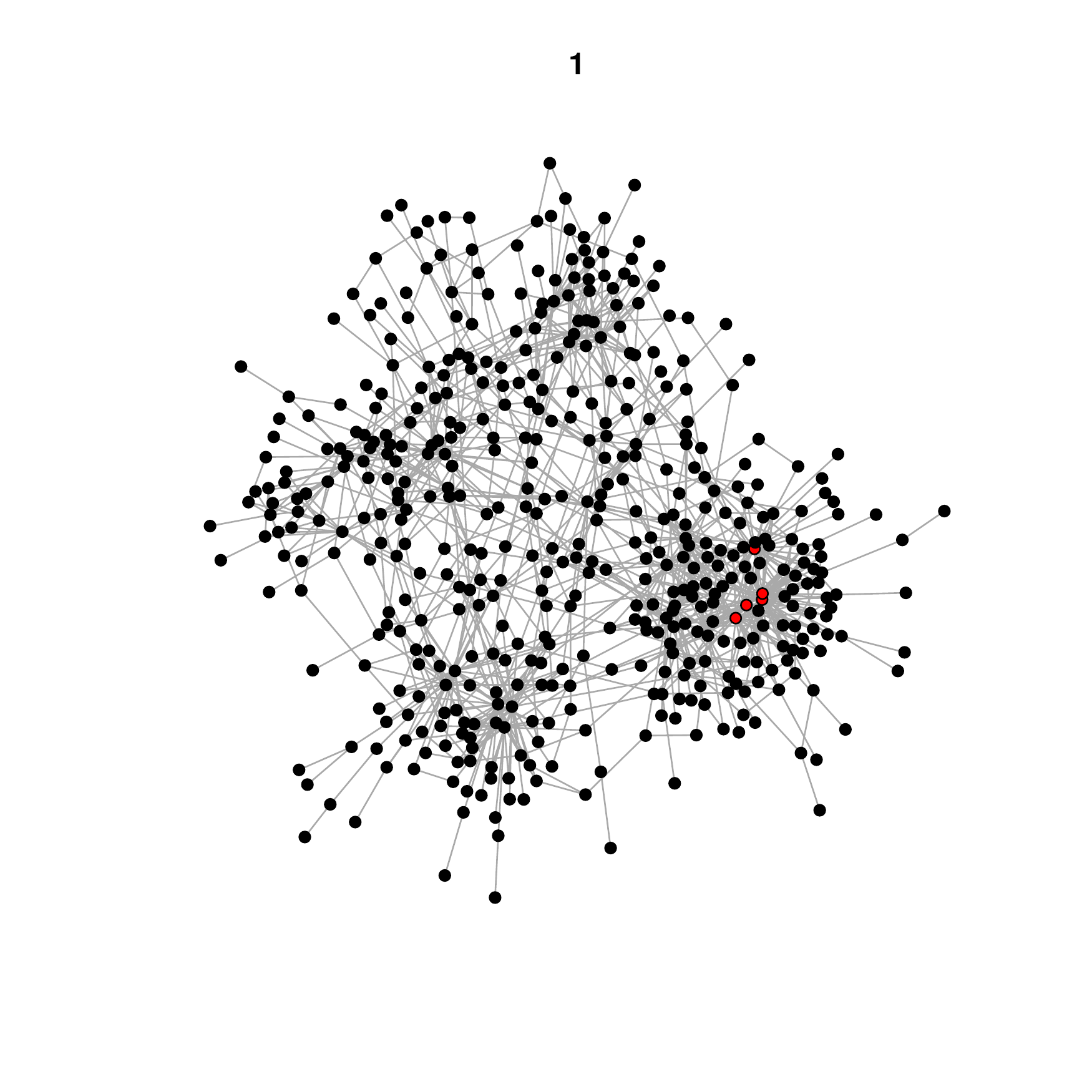}}&
\subfloat[p=500, power-law]{\includegraphics[height=3cm,width=5cm]{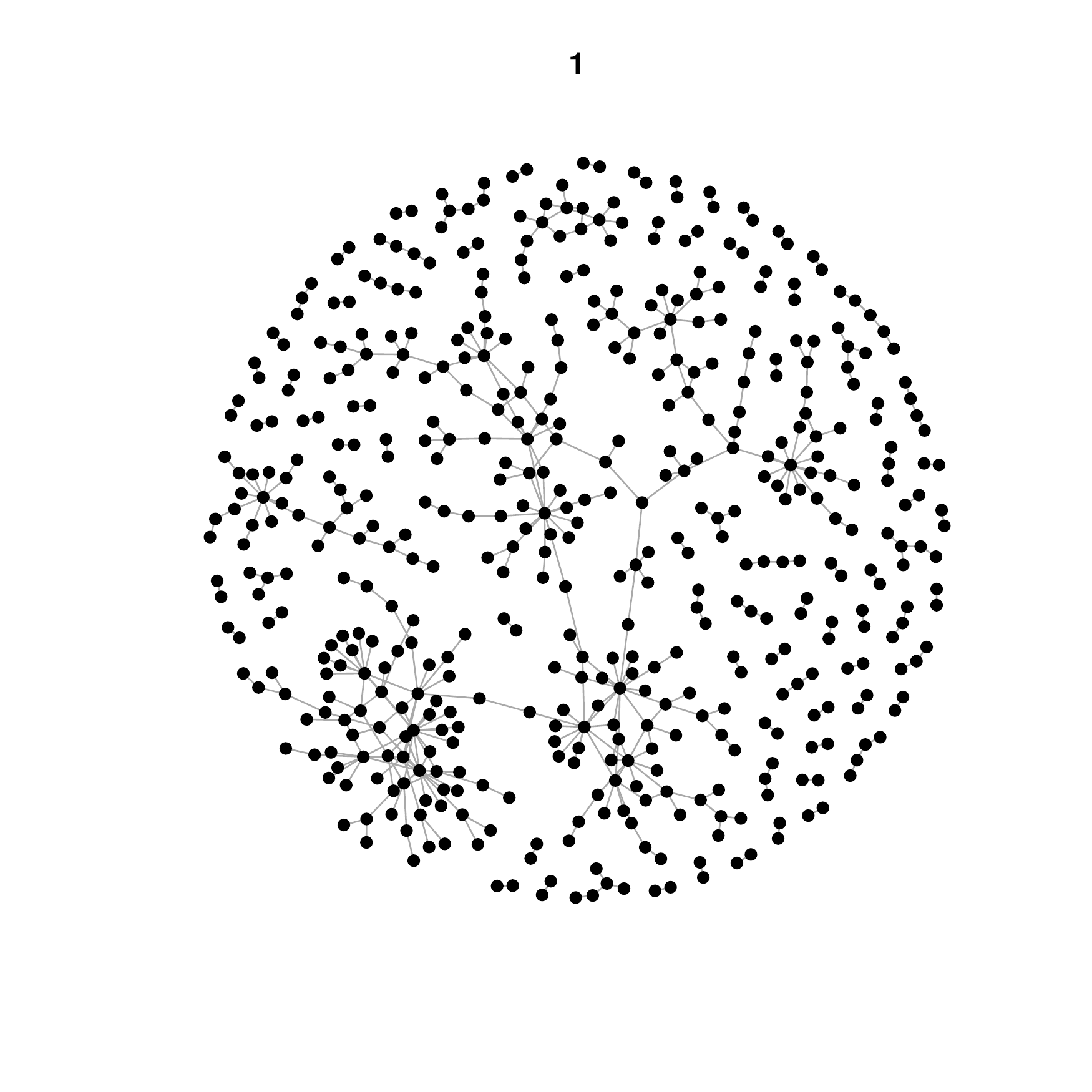}}\\
\end{tabular}
 \caption{Graph structure examples used in simulation data.}\label{fags}
\end{center}
\end{figure}

In Figure \ref{fags} we illustrate some of the simulated graph structures for all $p$ and the two models (hubs-based and power-law).

\subsection{Simulated data: regularization parameter selection}
In Table \ref{tabPOWER7} and Table \ref{tabPOWER1} we give the average ranks for the $\lambda$ selection magnitudes using the methods AGNES, A-MSE, PC, StARS, AIC (only GLasso) and BIC (only GLasso). The lowest rank (rank = 1) is assigned to the lowest choice of $\lambda$ and the largest rank (rank = 6 (GLasso), rank = 4 (MB)) is for the largest $\lambda$ out of all approaches. 

\begin{table}[h]
\tiny
\begin{center}
\caption{Average ranks for the $\lambda$ magnitude -GLasso estimates. } 
\begin{tabular}{ l r r r r  r  r r r r}
&\multicolumn{4}{c}{Hubs-based}&&\multicolumn{4}{c}{Power law}\\
\cline{2-5}\cline{7-10}
n & $50$ &$100$& $200$ & $500$ & &$50$ &$100$& $200$ & $500$\\
\hline
\multicolumn{10}{c}{dimension p=50} \\
AGNES  & 3.05 &3.55 &4.06 &4.40 &&3.12 &3.73 &4.40 &4.71\\
A-MSE& 4.33 &4.90 &5.22 &5.38 &&4.92 &5.47 &5.67 &5.78\\
PC     & 5.23 &5.80 &5.58 &5.15 &&4.58 &5.13 &4.85 &4.49\\
StARS  & 1.27 &1.49 &1.18 &1.28 &&1.17 &1.43 &1.02 &1.04\\
BIC    & 5.38 &3.73 &3.14 &3.06 &&5.33 &3.66 &3.08 &3.02\\
AIC    & 1.73 &1.52 &1.82 &1.73 &&1.90 &1.58 &1.98 &1.96\\
\multicolumn{10}{c}{dimension p=170} \\
AGNES  & 2.81 &3.30 &4.07 &4.63 &&2.69 &3.28 &3.91 &4.42\\
A-MSE& 4.13 &4.92 &5.17 &5.52 &&4.79 &5.38 &5.43 &5.63\\
PC     & 5.31 &5.91 &5.61 &4.03 &&4.96 &5.55 &5.46 &4.93\\
StARS  & 1.00 &1.22 &1.00 &1.00 &&1.00 &1.00 &1.00 &1.00\\
BIC    & 5.56 &3.87 &3.14 &3.48 &&5.23 &3.78 &3.20 &3.02\\
AIC    & 2.19 &1.78 &2.02 &2.35 &&2.33 &2.00 &2.00 &2.00\\
\multicolumn{10}{c}{dimension p=290} \\
AGNES  &2.54& 3.02& 3.97& 4.38&&2.33&3.07&3.87&4.20\\
A-MSE&4.17& 4.83& 5.08& 5.30&&4.83&5.39&5.46&5.46\\
PC     &5.12& 5.91& 5.78& 4.83&&4.68&5.57&5.53&5.33\\
StARS  &1.00& 1.01& 1.00& 1.00&&1.00&1.00&1.00&1.00\\
BIC    &5.71& 4.23& 3.13& 3.29&&5.47&3.98&3.14&3.02\\
AIC    &2.46& 1.99& 2.03& 2.20&&2.68&2.00&2.00&2.00\\
\multicolumn{10}{c}{dimension p=500} \\
AGNES  &2.13 &3.00&3.92& 4.25 &&2.11& 3.00& 3.62&4.11\\
A-MSE&4.28 &4.78&5.13& 5.36  &&4.81& 5.25& 5.27&5.47\\
PC     &4.94 &5.97&5.60& 5.72  &&4.63& 5.67& 5.73&5.39\\
StARS  &1.00 &1.00&1.00& 1.00  &&1.00& 1.00& 1.00&1.00\\
BIC    &5.78 &4.25&3.31& 3.89  &&5.55& 4.08& 3.38&3.03\\
AIC    &2.88 &2.00&2.05& 2.00  &&2.90& 2.00& 2.00&2.00\\
\end{tabular}
\label{tabPOWER7}
\end{center}
\end{table}
\vspace{-0.7cm} 
\singlespacing
\begin{table}[h]
\tiny
\begin{center}
\caption{Average ranks for the $\lambda$ magnitude -MB estimates. } 
\begin{tabular}{ l r r r r  r  r r r r}
&\multicolumn{4}{c}{Hubs-based}&&\multicolumn{4}{c}{Power law}\\
\cline{2-5}\cline{7-10}
n & $50$ &$100$& $200$ & $500$ & &$50$ &$100$& $200$ & $500$\\
\hline
\multicolumn{10}{c}{dimension p=50} \\
AGNES  &2.04& 2.08& 2.18& 2.15&&2.10& 2.33& 2.38& 2.78\\
A-MSE  &3.20& 3.22& 3.50& 3.45&&3.64& 3.72& 3.69& 3.89\\
PC     &3.76& 3.71& 3.32& 3.40&&3.26& 2.96& 2.93& 2.33\\
StARS  &1.00& 1.00& 1.00& 1.00&&1.00& 1.00& 1.00& 1.00\\ 
\multicolumn{10}{c}{dimension p=170} \\
AGNES  &2.00& 2.00& 2.02& 2.13&&2.01& 2.02& 2.17& 2.42\\
A-MSE  &3.27& 3.05& 3.10& 3.26&&3.62& 3.52& 3.64& 3.74\\
PC     &3.73& 3.95& 3.88& 3.61&&3.38& 3.47& 3.18& 2.83\\
StARS  &1.00& 1.00& 1.00& 1.00&&1.00& 1.00& 1.00& 1.00\\  
\multicolumn{10}{c}{dimension p=290} \\
AGNES  &2.00&  2.0& 2.00& 2.05&&2.00& 2.00& 2.04& 2.15\\
A-MSE  &3.42&  3.1& 3.12& 3.26&&3.73& 3.69& 3.61& 3.63\\
PC     &3.58&  3.9& 3.88& 3.69&&3.27& 3.31& 3.35& 3.22\\
StARS  &1.00& 1.00& 1.00& 1.00&&1.00& 1.00& 1.00& 1.00\\  
\multicolumn{10}{c}{dimension p=500} \\
AGNES  &2.00& 2.00& 2.00& 2.07&&2.00& 2.00& 2.00& 2.06\\
A-MSE  &3.55& 3.09& 3.04& 3.21&&3.76& 3.62& 3.54& 3.63\\
PC     &3.45& 3.91& 3.96& 3.73&&3.24& 3.38& 3.46& 3.31\\
StARS  &1.00& 1.00& 1.00& 1.00&&1.00& 1.00& 1.00& 1.00\\  
\end{tabular}
\label{tabPOWER1}
\end{center}
\end{table}
\onehalfspacing

\subsection{Simulated data: MSE ranks for network statistics}
We compute four basic global network statistics to compare the performance of the methods in terms of structure similarities with the theoretical graphs:
\begin{itemize}
\item The \emph{harmonic mean} of the geodesic distances in a graph is given by
\begin{equation}\label{eq:PC1}
H(\lambda)^{-1} = \left(\frac{2}{p(p-1)} \sum\limits_{i<j} \frac{1}{\hat{g}_{ij}(\lambda)}\right)^{-1}
\end{equation}
\item The AGNES coefficient is defined in Section 3.3 in the article and aims to measure the clustering structure of the network
\item The \emph{Estrada index}  quantifies the complexity of the network in terms of subgraphs and can be expressed for regular networks by
\begin{equation}
EE(\lambda)  = \sum_{j=1}^p \exp(\gamma_j(\lambda)),
\end{equation}\label{eq:EE1}
where $\gamma_1(\lambda), \ldots,\gamma_p(\lambda)$ are the eigenvalues of $\hat{A}_G^\lambda$. 
\item The square average dissimilarity matrix is given by
$$
AD(\lambda) = 2/[p(p-1)]\sum_{i<j} \hat{d}_{ij}(\lambda) . 
$$
\end{itemize}
Moreover, we also use the average MSE of the dissimilarity matrix and the degree distribution of the estimated networks. From Table 3-7 and Table 8-13 we present the loss function ranks for the square differences between the estimated graph structures and the true network using GLasso estimates and MB estimates respectively. The lowest rank (rank = 1) is assigned to the lowest square error and the largest rank (rank = 6 (GLasso), rank = 4 (MB)) is for the largest square error out of all approaches. 

\singlespacing
\begin{table}[h]
\tiny
\begin{center}
\caption{Average ranks for the mean square errors of the harmonic mean -GLasso estimates. } 
\begin{tabular}{ l r r r r  r  r r r r}
&\multicolumn{4}{c}{Hubs-based}&&\multicolumn{4}{c}{Power law}\\
\cline{2-5}\cline{7-10}
n & $50$ &$100$& $200$ & $500$ & &$50$ &$100$& $200$ & $500$\\
\hline
\multicolumn{10}{c}{dimension p=50} \\
AGNES  &\textbf{2.07}& 2.47& 2.23& 2.16&&\textbf{2.23}& 2.77& 2.12& 2.11\\
A-MSE&2.48& \textbf{2.17}& \textbf{1.62}& \textbf{1.39}&&3.10& \textbf{2.02}& \textbf{1.47}& \textbf{1.32}\\
PC     &4.47& 4.72& 3.88& 3.40&&3.09& 2.57& 3.27& 2.71\\
StARS  &3.88& 4.54& 5.34& 5.45&&4.58& 5.31& 5.74& 5.92\\
BIC    &4.68& 2.60& 3.24& 3.59&&4.11& 3.19& 3.62& 3.92\\
AIC    &3.42& 4.51& 4.69& 5.01&&3.88& 5.15& 4.79& 5.01\\
\multicolumn{10}{c}{dimension p=170} \\
AGNES  &\textbf{1.36}& \textbf{2.10}& 1.87& 1.87&&\textbf{1.89}& 2.37& \textbf{2.19}& 1.98\\
A-MSE&3.62& 2.42& \textbf{1.57}& \textbf{1.27}&&3.86& 3.15& 2.47& \textbf{1.55}\\
PC     &5.28& 5.59& 5.19& 4.71&&4.33& 4.07& 3.39& 3.17\\
StARS  &3.25& 4.60& 5.23& 5.67&&3.80& 5.10& 5.47& 5.80\\
BIC    &5.49& 2.25& 2.94& 3.17&&4.78& \textbf{2.27}& 3.07& 3.72\\
AIC    &2.01& 4.03& 4.20& 4.32&&2.34& 4.05& 4.42& 4.78\\
\multicolumn{10}{c}{dimension p=290} \\
AGNES  &\textbf{1.39} &\textbf{1.82} &\textbf{1.65} &1.78&&\textbf{1.82}& 2.23& \textbf{1.95}& 1.93\\
A-MSE&4.03 &3.27 &2.03 &\textbf{1.42}&&4.35& 4.14& 3.12& \textbf{1.69}\\
PC     &5.12 &5.59 &5.62 &4.88&&4.27& 4.68& 3.97& 3.52\\
StARS  &2.93 &4.41 &5.07 &5.50&&3.42& 4.48& 5.12& 5.67\\
BIC    &5.71 &2.53 &2.62 &3.14&&5.28& \textbf{2.06}& 2.74& 3.53\\
AIC    &1.81 &3.38 &4.02 &4.28&&\textbf{1.87}& 3.40& 4.10& 4.65\\
\multicolumn{10}{c}{dimension p=500} \\
AGNES  &\textbf{1.32}& \textbf{1.27}&\textbf{1.66}&\textbf{1.65} &&\textbf{1.57}& \textbf{1.85}& \textbf{1.89}&\textbf{1.72}\\
A-MSE&4.28& 4.20&2.78&1.88 &&4.76& 4.40& 3.18&2.14\\
PC     &4.94& 5.97&5.47&5.17 &&4.53& 5.42& 4.88&4.06\\
StARS  &2.48& 3.75&4.96&5.35 &&2.92& 4.12& 4.93&5.39\\
BIC    &5.78& 3.18&2.25&2.82 &&5.55& 2.17& 2.24&3.31\\
AIC    &2.21& 2.63&3.88&4.13 &&1.68& 3.05& 3.87&4.39\\
\end{tabular}
\label{tabPOWER5}
\end{center}
\end{table}

\vspace{-0.7cm }
\singlespacing
\begin{table}[H]
\tiny
\begin{center}
\caption{Average ranks for the mean square errors of the AGNES coefficient -GLasso estimates. } 
\begin{tabular}{ l r r r r  r  r r r r}
&\multicolumn{4}{c}{Hubs-based}&&\multicolumn{4}{c}{Power law}\\
\cline{2-5}\cline{7-10}
n & $50$ &$100$& $200$ & $500$ & &$50$ &$100$& $200$ & $500$\\
\hline
\multicolumn{10}{c}{dimension p=50} \\
AGNES  &2.77 &3.03& 2.52 &2.31&&\textbf{2.35}& 2.58& 2.18& 2.12\\
A-MSE&\textbf{2.23} &2.22& 1.90 &\textbf{1.68}&&2.82& 2.48& \textbf{2.07}& \textbf{1.74}\\
PC     &2.75 &\textbf{2.03} &\textbf{1.83} &2.23&&2.41& \textbf{2.13}& 2.15& 2.17\\
StARS  &5.46 &5.41 &5.33 &5.37&&5.62& 5.49& 5.33& 5.24\\
BIC    &2.60 &2.93 &3.98 &4.03&&2.98& 2.98& 3.62& 3.99\\
AIC    &5.19 &5.38 &5.44 &5.39&&4.83& 5.33& 5.66& 5.72\\
\multicolumn{10}{c}{dimension p=170} \\
AGNES  &2.98& 3.30& 2.65& 2.13&&\textbf{2.33}& 2.80& 2.27& 2.27\\
A-MSE&\textbf{1.95}& \textbf{1.94}& \textbf{1.70}&\textbf{1.50}&&2.83& 2.45& 2.33& \textbf{1.68}\\
PC     &2.52& 1.96& 1.84& 3.38&&3.14& \textbf{2.37}& \textbf{1.92}& 2.22\\
StARS  &5.98& 5.78& 4.78& 3.83&&6.00& 6.00& 4.78& 4.45\\
BIC    &3.26& 2.80& 4.03& 4.16&&3.50& 2.38& 3.68& 4.38\\
AIC    &4.31& 5.22& 6.00& 6.00&&3.21& 5.00& 6.00& 6.00\\
\multicolumn{10}{c}{dimension p=290} \\
AGNES  &3.08& 3.58& 2.65& 2.13&&\textbf{2.42}& 2.78& 2.57& 2.52\\
A-MSE&\textbf{2.13}& 2.17& \textbf{1.93}& \textbf{1.95}&&3.23& 2.73& \textbf{1.88}& \textbf{1.89}\\
PC     &2.86& \textbf{1.84}& 2.35& 2.88&&3.00& 2.63& 2.35& 2.04\\
StARS  &6.00& 5.99& 4.27& 3.50&&6.00& 6.00& 4.48& 3.87\\
BIC    &3.56& 2.42& 3.80& 4.54&&3.88& \textbf{1.86}& 3.73& 4.68\\
AIC    &3.38& 5.01& 6.00& 6.00&&2.47& 5.00& 6.00& 6.00\\
\multicolumn{10}{c}{dimension p=500} \\
AGNES  &2.83& 3.53&2.82&2.37 &&2.33& 3.03& 3.29&2.67\\
A-MSE&2.69& \textbf{2.08}&\textbf{2.04}&\textbf{2.02} &&3.64& 2.33& \textbf{2.13}&\textbf{2.03}\\
PC     &3.44& 2.13&2.42&3.37 &&3.18& 2.53& 2.15&2.28\\
StARS  &6.00& 6.00&4.02&3.60 &&6.00& 6.00& 3.65&3.22\\
BIC    &3.96& 2.30&3.73&3.80 &&4.20& \textbf{2.10}& 3.77&4.81\\
AIC    &\textbf{2.08}& 4.95&5.97&5.85 &&\textbf{1.65}& 5.00& 6.00&6.00\\
\end{tabular}
\label{tabPOWER6}
\end{center}
\end{table}

\begin{table}[H]
\tiny
\begin{center}
\caption{Average ranks for the mean square errors of the Estrada index -GLasso estimates. } 
\begin{tabular}{ l r r r r  r  r r r r}
&\multicolumn{4}{c}{Hubs-based}&&\multicolumn{4}{c}{Power law}\\
\cline{2-5}\cline{7-10}
n & $50$ &$100$& $200$ & $500$ & &$50$ &$100$& $200$ & $500$\\
\hline
\multicolumn{10}{c}{dimension p=50} \\
AGNES  & 3.70 &3.43 &2.96 &2.56 &&3.82& 3.28& 2.58 &2.29\\
A-MSE& 2.25 &2.02 &1.77 &\textbf{1.66} &&2.10& \textbf{1.52}& \textbf{1.33} &\textbf{1.21}\\
PC     & 2.10 &\textbf{1.35} &\textbf{1.45} &1.87 &&2.31& 1.92& 2.22 &2.52\\
StARS  & 5.72 &5.51 &5.83 &5.72 &&5.83& 5.58& 5.97 &5.96\\
BIC    & \textbf{2.02} &3.22 &3.83 &3.92 &&\textbf{1.88}& 3.29& 3.87 &3.98\\
AIC    & 5.21 &5.47 &5.17 &5.28 &&5.07& 5.42& 5.03 &5.04\\
\multicolumn{10}{c}{dimension p=170} \\
AGNES  &4.03 &3.68 &2.93 &2.37 &&4.31 &3.73& 3.09& 2.58\\
A-MSE&2.48 &2.09 &1.83 &\textbf{1.57} &&2.14 &1.63& 1.60& \textbf{1.40}\\
PC     &1.99 &\textbf{1.14} &\textbf{1.41} &2.96 &&2.04 &\textbf{1.43}& \textbf{1.52}& 2.13\\
StARS  &6.00 &5.78 &6.00 &6.00 &&6.00 &6.00& 6.00& 6.00\\
BIC    &\textbf{1.82} &3.08 &3.84 &3.46 &&\textbf{1.88} &3.20& 3.78& 3.88\\
AIC    &4.67 &5.22 &4.98 &4.65 &&4.62 &5.00& 5.00& 5.00\\
\multicolumn{10}{c}{dimension p=290} \\
AGNES  &3.84& 3.83& 2.97& 2.57&&4.65& 3.93& 3.13& 2.88\\
A-MSE&2.65& 2.12& 2.03& \textbf{1.87}&&2.15& 1.61& 1.61& \textbf{1.69}\\
PC     &2.39& \textbf{1.44}& \textbf{1.25}& 2.24&&2.25& \textbf{1.43}& \textbf{1.47}& 1.71\\
StARS  &6.00& 5.99& 6.00& 6.00&&6.00& 6.00& 6.00& 6.00\\
BIC    &\textbf{2.19}& 2.62& 3.78& 3.52&&\textbf{1.70}& 3.04& 3.79& 3.72\\
AIC    &3.92& 5.01& 4.97& 4.80&&4.25& 4.98& 5.00& 5.00\\
\multicolumn{10}{c}{dimension p=500} \\
AGNES  &4.42& 3.90&3.14& 2.50&&4.84& 3.90& 3.29&2.83\\
A-MSE&2.67& 2.15&1.98& 1.69&&2.19& 1.88& 1.93&\textbf{1.75}\\
PC     &2.29& \textbf{1.37}&\textbf{1.42}& \textbf{1.47}&&2.37& \textbf{1.47}& \textbf{1.50}&1.94\\
StARS  &6.00& 6.00&6.00& 6.00&&6.00& 6.00& 6.00&6.00\\
BIC    &\textbf{2.01}& 2.72&3.51& 4.71&&\textbf{1.57}& 2.98& 3.36&3.47\\
AIC    &3.61& 4.87&4.95& 5.00&&4.03& 4.77& 4.92&5.00\\
\end{tabular}
\label{tabPOWER4}
\end{center}
\end{table}

\begin{table}[H]
\tiny
\begin{center}
\caption{Average ranks for the mean square errors of the average dissimilarity matrix -GLasso estimates.} 
\begin{tabular}{ l r r r r  r  r r r r}
&\multicolumn{4}{c}{Hubs-based}&&\multicolumn{4}{c}{Power law}\\
\cline{2-5}\cline{7-10}
n & $50$ &$100$& $200$ & $500$ & &$50$ &$100$& $200$ & $500$\\
\hline
\multicolumn{10}{c}{dimension p=50} \\
AGNES  &\textbf{2.17}& 2.75& 2.34& 2.19&&\textbf{2.43} &2.67 &2.38 &2.11\\
A-MSE&2.75& \textbf{2.68}& \textbf{1.95}& \textbf{1.69}&&2.52 &\textbf{1.78} &\textbf{1.63} &\textbf{1.52}\\
PC     &4.05& 4.00& 2.65& 2.55&&2.58 &2.25 &2.32 &2.48\\
StARS  &4.12& 4.46& 5.81& 5.72&&5.68 &5.53 &5.97 &5.96\\
BIC    &4.28& \textbf{2.68}& 3.12& 3.58&&3.26 &3.42 &3.67 &3.89\\
AIC    &3.62& 4.42& 5.12& 5.28&&4.53 &5.35 &5.03 &5.04\\
\multicolumn{10}{c}{dimension p=170} \\
AGNES  &\textbf{1.61}& 2.43& 2.15& 2.03&&\textbf{2.59}& 2.80& 2.84& 2.53\\
A-MSE&2.73& \textbf{1.94}& \textbf{1.78}& \textbf{1.25}&&2.76& \textbf{2.22}& \textbf{1.72}& \textbf{1.38}\\
PC     &3.98& 3.16& 3.06& 3.59&&3.01& 2.43& 1.74& 2.10\\
StARS  &6.00& 5.78& 6.00& 6.00&&6.00& 6.00& 6.00& 6.00\\
BIC    &4.29& 2.48& 3.02& 3.48&&3.45& 2.55& 3.70& 3.98\\
AIC    &2.39& 5.20& 4.98& 4.65&&3.19& 5.00& 5.00& 5.00\\
\multicolumn{10}{c}{dimension p=290} \\
AGNES  &\textbf{1.57}& \textbf{2.11}& 2.27& 2.33&&\textbf{2.47}& 2.85& 3.07& 2.78\\
A-MSE&3.00& 2.48& \textbf{1.98}& \textbf{1.55}&&3.15& 2.46& \textbf{1.48}& \textbf{1.57}\\
PC     &4.06& 3.64& 2.85& 2.67&&3.12& 2.50& 1.72& 1.66\\
StARS  &6.00& 5.99& 6.00& 6.00&&6.00& 6.00& 6.00& 6.00\\
BIC    &4.64& 2.32& 2.93& 3.64&&3.85& \textbf{2.23}& 3.74& 3.98\\
AIC    &1.73& 4.46& 4.97& 4.80&&2.42& 4.97& 5.00& 5.00\\
\multicolumn{10}{c}{dimension p=500} \\
AGNES  &\textbf{1.22}& \textbf{1.75}&2.10& 1.68&&\textbf{1.82}& 3.28& 3.23&2.89\\
A-MSE&3.26& 2.93&\textbf{2.09}& \textbf{1.67}&&3.46& 2.27& 1.67&\textbf{1.53}\\
PC     &3.92& 4.28&3.32& 3.15&&3.33& 2.62& \textbf{1.62}&1.61\\
StARS  &6.00& 6.00&6.00& 6.00&&6.00& 6.00& 6.00&6.00\\
BIC    &4.78& 2.52&2.55& 3.39&&4.30& \textbf{1.92}& 3.49&3.97\\
AIC    &1.82& 3.52&4.94& 5.00&&2.09& 4.92& 5.00&5.00\\
\end{tabular}
\label{tabPOWER2}
\end{center}
\end{table}

\begin{table}[h]
\tiny
\begin{center}
\caption{Average ranks for the absolute error on the degree distribution -GLasso estimates. }  
\begin{tabular}{ l r r r r  r  r r r r}
&\multicolumn{4}{c}{Hubs-based}&&\multicolumn{4}{c}{Power law}\\
\cline{2-5}\cline{7-10}
n & $50$ &$100$& $200$ & $500$ & &$50$ &$100$& $200$ & $500$\\
\hline
\multicolumn{10}{c}{dimension p=50} \\
AGNES&3.00 &3.09 &2.50 &2.54 &&3.69 &2.98 &2.62 &2.19\\
A-MSE&\textbf{2.02} &\textbf{1.77} &\textbf{1.51} &\textbf{1.55} &&2.22 &\textbf{1.56} &\textbf{1.36} &\textbf{1.27}\\
PC   &2.51 &1.95 &2.06 &2.05 &&\textbf{2.08} &2.23 &2.16 &2.54\\
StARS&5.72 &5.52 &5.78 &5.72 &&5.88 &5.57 &5.97 &5.89\\
BIC  &2.48 &3.20 &3.93 &3.87 &&2.10 &3.24 &3.86 &3.99\\
AIC  &5.28 &5.47 &5.22 &5.27 &&5.04 &5.42 &5.03 &5.11\\
\multicolumn{10}{c}{dimension p=170} \\
AGNES&3.08 &3.38 &2.92 &2.58 &&4.23 &3.81 &3.03 &2.74\\
A-MSE&\textbf{1.99} &\textbf{1.76} &1.67 &\textbf{1.59} &&2.11 &1.91 &1.67 &\textbf{1.44}\\
PC   &2.73 &2.20 &\textbf{1.53} &2.27 &&\textbf{1.82} &\textbf{1.61} &\textbf{1.42} &1.82\\
StARS&6.00 &5.86 &6.00 &6.00 &&6.00 &6.00 &6.00 &6.00\\
BIC  &2.94 &2.67 &3.90 &3.72 &&2.20 &2.67 &3.88 &4.00\\
AIC  &4.25 &5.14 &4.98 &4.83 &&4.63 &5.00 &5.00 &5.00\\
\multicolumn{10}{c}{dimension p=290} \\
AGNES&3.00 &3.55 &2.73 &2.52 &&4.48 &3.83 &3.11 &2.83\\
A-MSE&\textbf{2.09} &\textbf{1.83} &\textbf{1.67} &\textbf{1.53} &&\textbf{2.11} &1.77 &1.62 &\textbf{1.44}\\
PC   &3.15 &2.26 &1.73 &2.40 &&2.13 &\textbf{1.59} &\textbf{1.41} &1.75\\
StARS&6.00 &6.00 &6.00 &6.00 &&6.00 &6.00 &6.00 &6.00\\
BIC  &3.52 &2.36 &3.88 &3.68 &&2.28 &2.80 &3.87 &3.98\\
AIC  &3.23 &5.00 &5.00 &4.87 &&3.99 &5.00 &5.00 &5.00\\
\multicolumn{10}{c}{dimension p=500} \\
AGNES&2.65& 3.00& 2.84& 2.54 &&4.72 &3.95 &3.31 &2.85\\
A-MSE&2.80& 1.88& 1.72& 1.55 &&2.59 &1.88 &1.62 &\textbf{1.57}\\
PC   &3.27& 2.25& 1.78& 2.62 &&\textbf{1.94} &\textbf{1.64} &\textbf{1.41} &1.60\\
StARS&6.00& 6.00 &6.00 &6.00 &&6.00 &6.00 &6.00 &6.00\\
BIC  &4.19& 2.00& 3.90& 3.55 &&2.71 &2.53 &3.66 &3.98\\
AIC  &2.08& 4.64& 5.00& 4.75 &&3.04 &5.00 &5.00 &5.00\\
\end{tabular}
\label{tabPOWER3}
\end{center}
\end{table}

\begin{table}[H]
\tiny
\begin{center}
\caption{Average ranks for the mean square errors of the harmonic mean -MB estimates. } 
\begin{tabular}{ l r r r r  r  r r r r}
&\multicolumn{4}{c}{Hubs-based}&&\multicolumn{4}{c}{Power law}\\
\cline{2-5}\cline{7-10}
n & $50$ &$100$& $200$ & $500$ & &$50$ &$100$& $200$ & $500$\\
\hline
\multicolumn{10}{c}{dimension p=50} \\
AGNES  &\textbf{1.86}& 1.99& 2.43& 2.67&&\textbf{1.95}& \textbf{1.81}& 1.95& 1.86\\
A-MSE  &1.87& \textbf{1.48}& \textbf{1.32}& \textbf{1.37}&&2.26& 2.31& \textbf{1.77}& \textbf{1.66}\\
PC     &3.12& 3.23& 2.50& 2.07&&2.29& 2.25& 2.47& 2.53\\
StARS  &3.15& 3.30& 3.75& 3.90&&3.50& 3.63& 3.82& 3.95\\
\multicolumn{10}{c}{dimension p=170} \\
AGNES  &\textbf{1.53}& 2.18& 2.40& 2.47&&\textbf{1.39}& \textbf{1.60}& \textbf{1.82}& \textbf{1.61}\\
A-MSE  &2.31& \textbf{1.43}& \textbf{1.33}& \textbf{1.27}&&3.08& 2.65& 2.54& 2.26\\
PC     &3.42& 2.98& 2.62& 2.58&&2.88& 2.83& 2.20& 2.32\\
StARS  &2.73& 3.40& 3.65& 3.68&&2.65& 2.92& 3.43& 3.82\\
\multicolumn{10}{c}{dimension p=290} \\
AGNES  &\textbf{1.20}& 1.98& 2.20& 2.53&&\textbf{1.20}& \textbf{1.40}& \textbf{1.41}& \textbf{1.55}\\
A-MSE  &3.12& \textbf{1.53}& \textbf{1.41}& \textbf{1.24}&&3.52& 3.23& 2.91& 2.47\\
PC     &3.35& 3.37& 2.92& 2.51&&3.02& 2.73& 2.65& 2.35\\
StARS  &2.33& 3.12& 3.47& 3.72&&2.27& 2.65& 3.03& 3.63\\
\multicolumn{10}{c}{dimension p=500} \\
AGNES  &\textbf{1.07}& \textbf{1.65}& 2.07& 2.43&&\textbf{1.05}& \textbf{1.25}& \textbf{1.72}& \textbf{1.43}\\
A-MSE  &3.53& 1.96& \textbf{1.32}& \textbf{1.32}&&3.74& 3.38& 2.77& 2.78\\
PC     &3.30& 3.49& 3.34& 2.62&&3.14& 3.08& 2.62& 2.54\\
StARS  &2.10& 2.90& 3.27& 3.62&&2.07& 2.28& 2.88& 3.25\\
\end{tabular}
\label{tabPOWER8}
\end{center}
\end{table}

\begin{table}[H]
\tiny
\begin{center}
\caption{Average ranks for the mean square errors of the AGNES coefficient -MB estimates. } 
\begin{tabular}{ l r r r r  r  r r r r}
&\multicolumn{4}{c}{Hubs-based}&&\multicolumn{4}{c}{Power law}\\
\cline{2-5}\cline{7-10}
n & $50$ &$100$& $200$ & $500$ & &$50$ &$100$& $200$ & $500$\\
\hline
\multicolumn{10}{c}{dimension p=50} \\
AGNES  &2.46& 2.77& 2.70& 2.82&&2.00& 2.33& 2.58& 2.20\\
A-MSE  &1.90& 1.72& \textbf{1.32}& \textbf{1.48}&&2.26& 1.90& \textbf{1.61}& \textbf{1.48}\\
PC     &\textbf{1.64}& \textbf{1.51}& 2.00& 1.90&&\textbf{1.76}& \textbf{1.77}& 1.87& 2.38\\
StARS  &4.00& 4.00& 3.98& 3.80&&3.98& 4.00& 3.95& 3.93\\
\multicolumn{10}{c}{dimension p=170} \\
AGNES  &2.67& 3.00& 3.22& 3.12&&2.39& 2.62& 2.66& 2.49\\
A-MSE  &1.71& 1.83& 1.83& 1.71&&1.87& 1.78& \textbf{1.68}& \textbf{1.66}\\
PC     &\textbf{1.62}& \textbf{1.17}& \textbf{1.25}& \textbf{1.64}&&\textbf{1.74}& \textbf{1.60}& 1.88& 2.18\\
StARS  &4.00& 4.00& 3.70& 3.53&&4.00& 4.00& 3.78& 3.67\\
\multicolumn{10}{c}{dimension p=290} \\
AGNES  &2.53& 2.97& 3.13& 3.47&&2.45& 2.85& 3.01& 2.93\\
A-MSE  &1.80& 1.73& 1.77& 1.73&&1.95& 1.59& \textbf{1.61}& \textbf{1.53}\\
PC     &\textbf{1.67}& \textbf{1.30}& \textbf{1.29}& \textbf{1.59}&&\textbf{1.60}& \textbf{1.56}& 1.80& 2.12\\
StARS  &4.00& 4.00& 3.80& 3.22&&4.00& 4.00& 3.58& 3.42\\
\multicolumn{10}{c}{dimension p=500} \\
AGNES  &2.57& 2.97& 3.20& 3.63&&2.32& 2.92& 3.17& 3.01\\
A-MSE  &1.73& 1.64& 1.81& \textbf{1.73}&&2.11& 1.58& \textbf{1.56}& \textbf{1.77}\\
PC     &\textbf{1.70}& \textbf{1.39}& \textbf{1.19}& 1.74&&\textbf{1.57}& \textbf{1.50}& 1.57& 2.11\\
StARS  &4.00& 4.00& 3.80& 2.90&&4.00& 4.00& 3.70& 3.12\\
\end{tabular}
\label{tabPOWER9}
\end{center}
\end{table}
\vspace{-0.7cm }
\begin{table}[H]
\tiny
\begin{center}
\caption{Average ranks for the mean square errors of the Estrada index -MB estimates. } 
\begin{tabular}{ l r r r r  r  r r r r}
&\multicolumn{4}{c}{Hubs-based}&&\multicolumn{4}{c}{Power law}\\
\cline{2-5}\cline{7-10}
n & $50$ &$100$& $200$ & $500$ & &$50$ &$100$& $200$ & $500$\\
\hline
\multicolumn{10}{c}{dimension p=50} \\
AGNES  &2.29& 2.56& 2.75& 2.82&&2.48& 2.44& 2.32& 1.92\\
A-MSE  &\textbf{1.67}& \textbf{1.62}& \textbf{1.33}& \textbf{1.47}&&1.79& \textbf{1.66}& \textbf{1.50}& \textbf{1.43}\\
PC     &2.04& 1.82& 1.92& 1.72&&\textbf{1.73}& 1.90& 2.18& 2.65\\
StARS  &4.00& 4.00& 4.00& 4.00&&4.00& 4.00& 4.00& 4.00\\
\multicolumn{10}{c}{dimension p=170} \\
AGNES  &\textbf{1.57}& 2.67& 2.93& 2.85&&2.08& 2.32& 2.33& 2.11\\
A-MSE  &1.98& \textbf{1.45}& \textbf{1.53}& 1.59&&2.02& \textbf{1.83}& \textbf{1.71}& \textbf{1.62}\\
PC     &2.46& 1.88& \textbf{1.53}& \textbf{1.56}&&\textbf{1.91}& 1.85& 1.97& 2.27\\
StARS  &4.00& 4.00& 4.00& 4.00&&4.00& 4.00& 4.00& 4.00\\
\multicolumn{10}{c}{dimension p=290} \\
AGNES  &\textbf{1.33}& 2.20& 2.70& 2.92&&\textbf{1.52}& 1.98& 2.34& 2.13\\
A-MSE  &2.23& \textbf{1.62}& \textbf{1.39}& \textbf{1.54}&&2.48& 2.17& 1.86& \textbf{1.77}\\
PC     &2.43& 2.18& 1.91& \textbf{1.54}&&2.00& \textbf{1.84}& \textbf{1.80}& 2.10\\
StARS  &4.00& 4.00& 4.00& 4.00&&4.00& 4.00& 4.00& 4.00\\
\multicolumn{10}{c}{dimension p=500} \\
AGNES  &\textbf{1.00}& \textbf{1.48}& 2.62& 2.90&&\textbf{1.22}& \textbf{1.83}& 2.32& 2.02\\
A-MSE  &2.55& 1.84& \textbf{1.31}& \textbf{1.49}&&2.66& 2.22& 1.88& \textbf{1.95}\\
PC     &2.45& 2.67& 2.08& 1.61&&2.12& 1.95& \textbf{1.81}& 2.02\\
StARS  &4.00& 4.00& 4.00& 4.00&&4.00& 4.00& 4.00& 4.00\\
\end{tabular}
\label{tabPOWER10}
\end{center}
\end{table}
\vspace{-0.7cm }
\begin{table}[H]
\tiny
\begin{center}
\caption{Average ranks for the mean square errors of the dissimilarity matrix -MB estimates. } 
\begin{tabular}{ l r r r r  r  r r r r}
&\multicolumn{4}{c}{Hubs-based}&&\multicolumn{4}{c}{Power law}\\
\cline{2-5}\cline{7-10}
n & $50$ &$100$& $200$ & $500$ & &$50$ &$100$& $200$ & $500$\\
\hline
\multicolumn{10}{c}{dimension p=50} \\
AGNES&2.21& 2.24& 2.52& 2.67&&2.53& 2.53& 2.36& 1.97\\
A-MSE&\textbf{1.82}& \textbf{1.43}& \textbf{1.30}& \textbf{1.42}&&\textbf{1.57}& \textbf{1.33}& \textbf{1.54}& \textbf{1.33}\\
PC   &2.49& 2.59& 2.18& 1.92&&1.94& 2.13& 2.10& 2.70\\
StARS&3.48& 3.73& 4.00& 4.00&&3.95& 4.00& 4.00& 4.00\\
\multicolumn{10}{c}{dimension p=170} \\
AGNES&2.87& 2.58& 2.65& 2.62&&2.94& 2.97& 2.54& 2.12\\
A-MSE&\textbf{1.51}& \textbf{1.42}& \textbf{1.40}& \textbf{1.29}&&\textbf{1.40}& \textbf{1.47}& \textbf{1.54}& \textbf{1.41}\\
PC   &1.62& 2.00& 1.95& 2.09&&1.66& 1.57& 1.92& 2.47\\
StARS&4.00& 4.00& 4.00& 4.00&&4.00& 4.00& 4.00& 4.00\\
\multicolumn{10}{c}{dimension p=290} \\
AGNES&2.88& 2.55& 2.60& 2.72&&3.0& 3.00& 2.84& 2.55\\
A-MSE&\textbf{1.52}& \textbf{1.4}0& \textbf{1.26}& \textbf{1.31}&&\textbf{1.3}& \textbf{1.39}& \textbf{1.43}& \textbf{1.47}\\
PC   &1.60& 2.05& 2.14& 1.98&&1.7& 1.61& 1.73& 1.98\\
StARS&4.00& 4.00& 4.00& 4.00&&4.0& 4.00& 4.00& 4.00\\
\multicolumn{10}{c}{dimension p=500} \\
AGNES  &2.90& 2.40& 2.45& 2.67&&3.00& 3.00& 3.00&2.34 \\
A-MSE&\textbf{1.47}& \textbf{1.27}& \textbf{1.18}& \textbf{1.34}&&\textbf{1.31}& \textbf{1.38}& \textbf{1.39}&\textbf{1.62} \\
PC     &1.63& 2.33& 2.38& 1.99&&1.69& 1.62& 1.61&2.04 \\
StARS  &4.00& 4.00& 4.00& 4.00&&4.00& 4.00& 4.00&4.00 \\
\end{tabular}
\label{tabPOWER11}
\end{center}
\end{table}
\vspace{-0.7cm }
\begin{table}[H]
\tiny
\begin{center}
\caption{Average ranks for the mean square errors of the average dissimilarity matrix -MB estimates } 
\begin{tabular}{ l r r r r  r  r r r r}
&\multicolumn{4}{c}{Hubs-based}&&\multicolumn{4}{c}{Power law}\\
\cline{2-5}\cline{7-10}
n & $50$ &$100$& $200$ & $500$ & &$50$ &$100$& $200$ & $500$\\
\hline
\multicolumn{10}{c}{dimension p=50} \\
AGNES  &\textbf{1.69}& \textbf{1.92}&  2.1& 2.27&&\textbf{1.85}& \textbf{1.97}& 1.98& 1.95\\
A-MSE&2.63& 2.20& \textbf{1.6}& \textbf{1.57}&&2.04& 2.00& \textbf{1.71}& \textbf{1.62}\\
PC     &3.17& 3.09&  2.3& 2.17&&2.38& 2.10& 2.32& 2.43\\
StARS  &2.50& 2.78&  4.0& 4.00&&3.73& 3.93& 4.00& 4.00\\
\multicolumn{10}{c}{dimension p=170} \\
AGNES  &\textbf{1.00}& \textbf{1.03}& \textbf{1.52}& 2.07&&\textbf{1.16}& \textbf{1.23}& \textbf{1.61}& \textbf{1.82}\\
A-MSE&2.29& 2.10& 1.87& \textbf{1.69}&&2.53& 2.38& 2.38& 1.96\\
PC     &2.71& 2.87& 2.62& 2.24&&2.31& 2.38& 2.02& 2.22\\
StARS  &4.00& 4.00& 4.00& 4.00&&4.00& 4.00& 4.00& 4.00\\
\multicolumn{10}{c}{dimension p=290} \\
AGNES  &\textbf{1.00}& \textbf{1.02}& \textbf{1.43}& 2.37&&\textbf{1.02}& \textbf{1.10}& \textbf{1.26}& \textbf{1.72}\\
A-MSE&2.42& 2.10& 1.94& \textbf{1.54}&&2.73& 2.64& 2.42& 2.20\\
PC     &2.58& 2.88& 2.62& 2.09&&2.25& 2.26& 2.32& 2.08\\
StARS  &4.00& 4.00& 4.00& 4.00&&4.00& 4.00& 4.00& 4.00\\
\multicolumn{10}{c}{dimension p=500} \\
AGNES  &\textbf{1.00}& \textbf{1.00}& \textbf{1.20}& 2.43&&\textbf{1.00}& \textbf{1.02}& \textbf{1.03}&\textbf{1.34}\\
A-MSE&2.55& 2.09& 1.94& \textbf{1.59}&&2.77& 2.55& 2.56&2.37\\
PC     &2.45& 2.91& 2.86& 1.98&&2.23& 2.43& 2.41&2.29\\
StARS  &4.00& 4.00& 4.00& 4.00&&4.00& 4.00& 4.00&4.00\\
\end{tabular}
\label{tabPOWER12}
\end{center}
\end{table}
\vspace{-0.7cm }
\begin{table}[H]
\tiny
\begin{center}
\caption{Average ranks for the mean square errors of the degree distribution  -MB estimates. } 
\begin{tabular}{ l r r r r  r  r r r r}
&\multicolumn{4}{c}{Hubs-based}&&\multicolumn{4}{c}{Power law}\\
\cline{2-5}\cline{7-10}
n & $50$ &$100$& $200$ & $500$ & &$50$ &$100$& $200$ & $500$\\
\hline
\multicolumn{10}{c}{dimension p=50} \\
AGNES  &2.39& 2.45& 2.61& 2.8&&2.30& 2.18& 2.19& 1.89\\
A-MSE  &\textbf{1.56}& \textbf{1.51}& \textbf{1.28}&\textbf{1.4}&&1.90& 1.93& \textbf{1.59}& \textbf{1.38}\\
PC     &2.05& 2.04& 2.11& 1.8&&\textbf{1.80}& \textbf{1.88}& 2.22& 2.73\\
StARS  &4.00& 4.00& 4.00& 4.0&&4.00& 4.00& 4.00& 4.00\\
\multicolumn{10}{c}{dimension p=170} \\
AGNES  &1.94& 2.72& 2.88& 2.84&&\textbf{1.76}& \textbf{1.82}& 1.92& \textbf{1.68}\\
A-MSE  &\textbf{1.71}& \textbf{1.35}& \textbf{1.41}& \textbf{1.38}&&2.26& 2.10& 2.15& 2.07\\
PC     &2.35& 1.93& 1.71& 1.78&&1.98& 2.08& \textbf{1.93}& 2.26\\
StARS  &4.00& 4.00& 4.00& 4.00&&4.00& 4.00& 4.00& 4.00\\
\multicolumn{10}{c}{dimension p=290} \\
AGNES  &\textbf{1.44}&  2.5& 2.85& 2.87&&\textbf{1.32}& \textbf{1.51}& \textbf{1.52}& \textbf{1.64}\\
A-MSE  &2.17&  \textbf{1.3}& \textbf{1.23}& \textbf{1.37}&&2.58& 2.48& 2.38& 2.33\\
PC     &2.39&  2.2& 1.92& 1.77&&2.10& 2.02& 2.10& 2.03\\
StARS  &4.00&  4.0& 4.00& 4.00&&4.00& 4.00& 4.00& 4.00\\
\multicolumn{10}{c}{dimension p=500} \\
AGNES  &\textbf{1.18}& 2.02& 2.73& 2.87&&\textbf{1.12}& \textbf{1.31}& \textbf{1.68}& \textbf{1.35}\\
A-MSE  &2.44& \textbf{1.48}& \textbf{1.07}& \textbf{1.31}&&2.71& 2.45& 2.23& 2.53\\
PC     &2.38& 2.51& 2.20& 1.82&&2.17& 2.24& 2.09& 2.12\\
StARS  &4.00& 4.00& 4.00& 4.00&&4.00& 4.00& 4.00& 4.00\\
\end{tabular}
\label{tabPOWER13}
\end{center}
\end{table}

\onehalfspacing

\newpage
\normalsize
\subsection{A-MSE and oracle tuning parameter}
\begin{figure}[H]
\begin{center}
 \begin{tabular}{cc}
\subfloat[p=50]{\includegraphics[width=6cm,height=4cm]{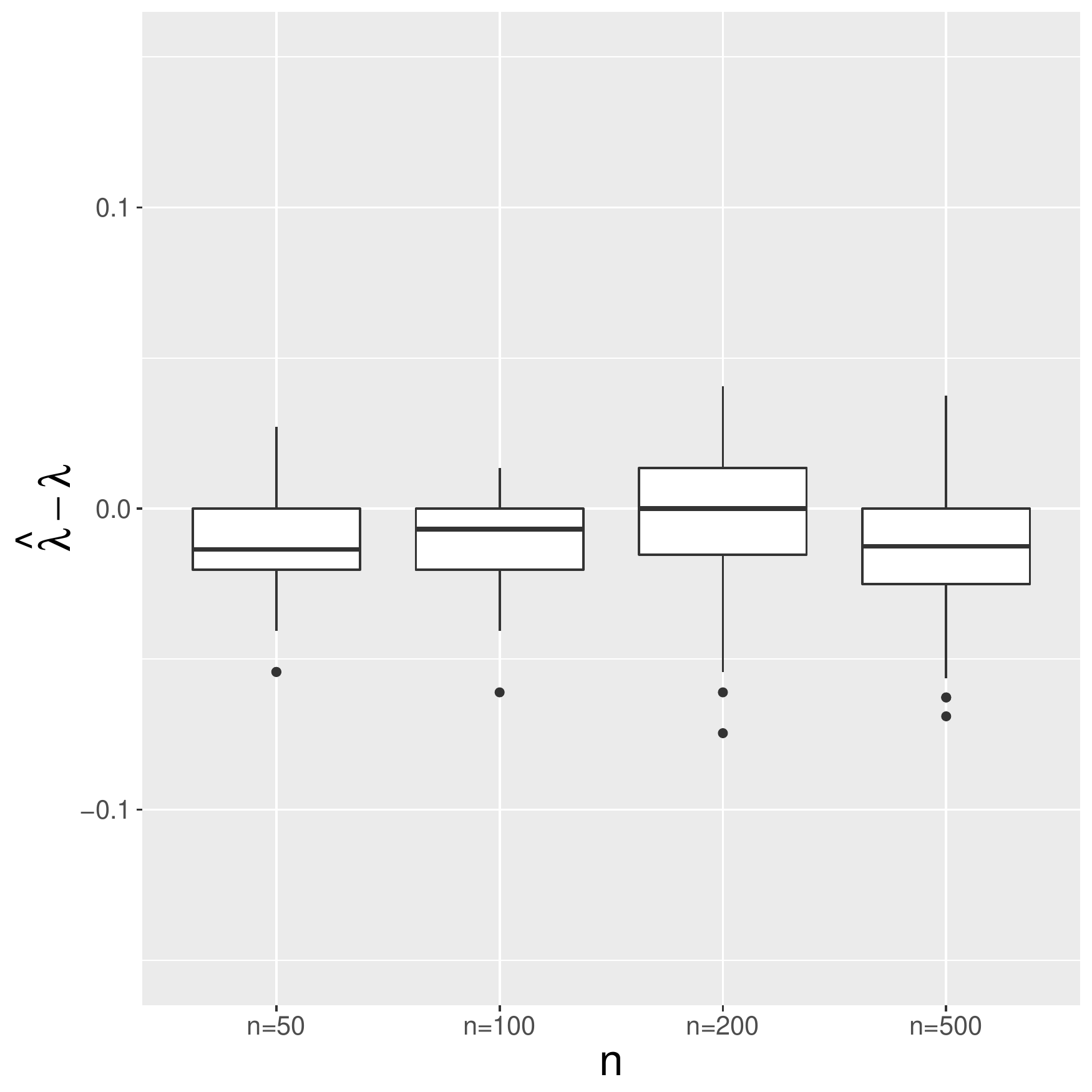}}&
\subfloat[p=170]{\includegraphics[width=6cm,height=4cm]{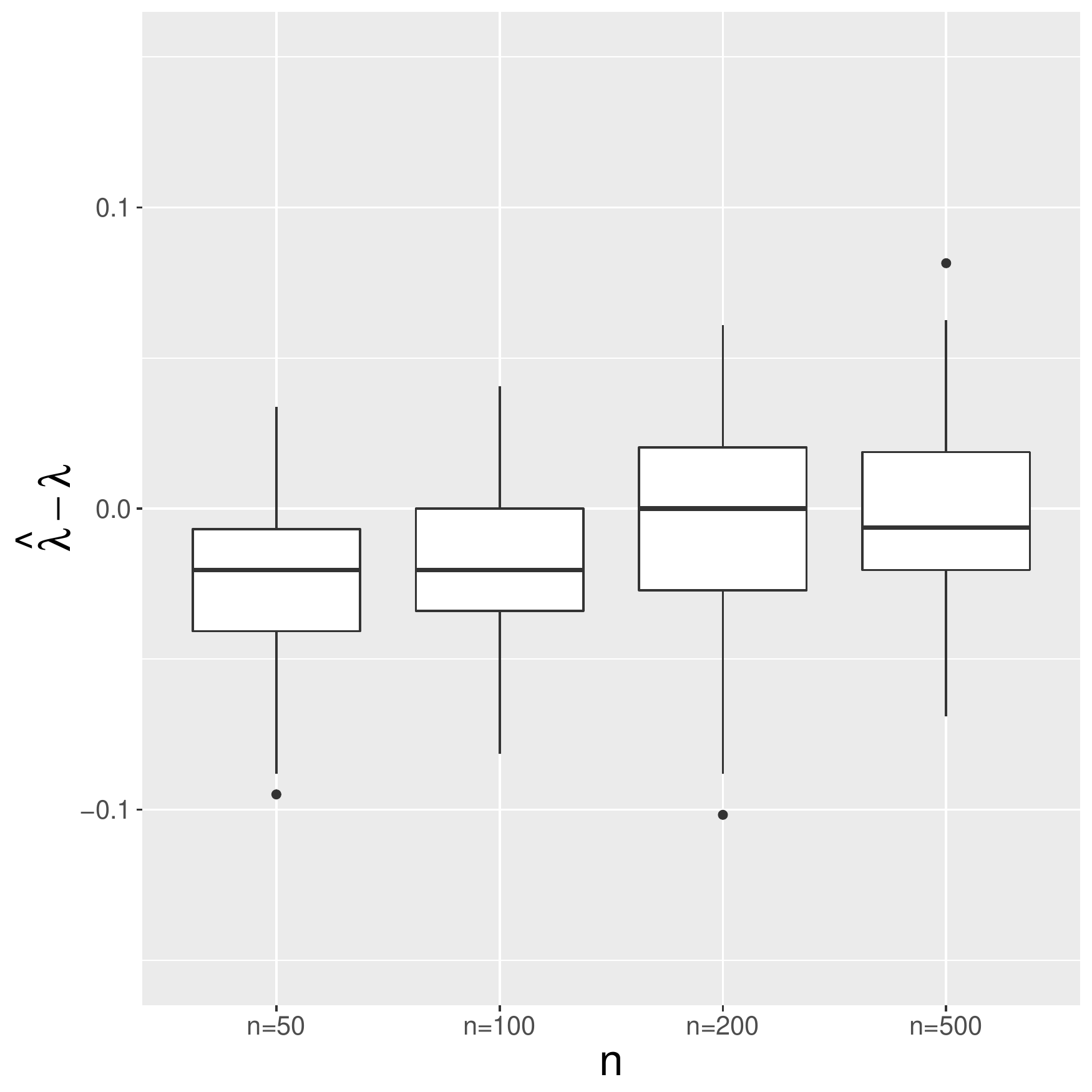}}\\
\subfloat[p=290]{\includegraphics[width=6cm,height=4cm]{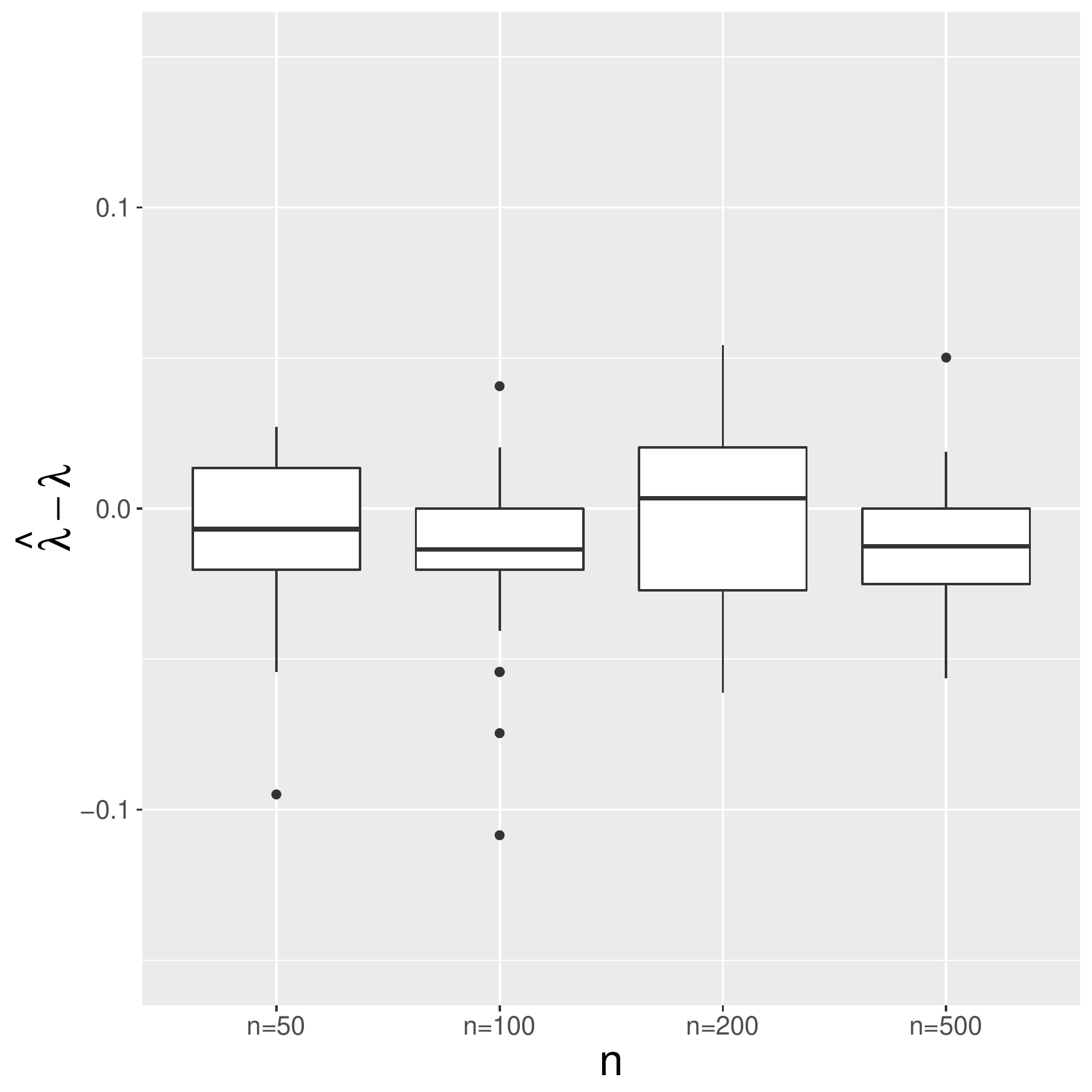}}&
\subfloat[p=500]{\includegraphics[width=6cm,height=4cm]{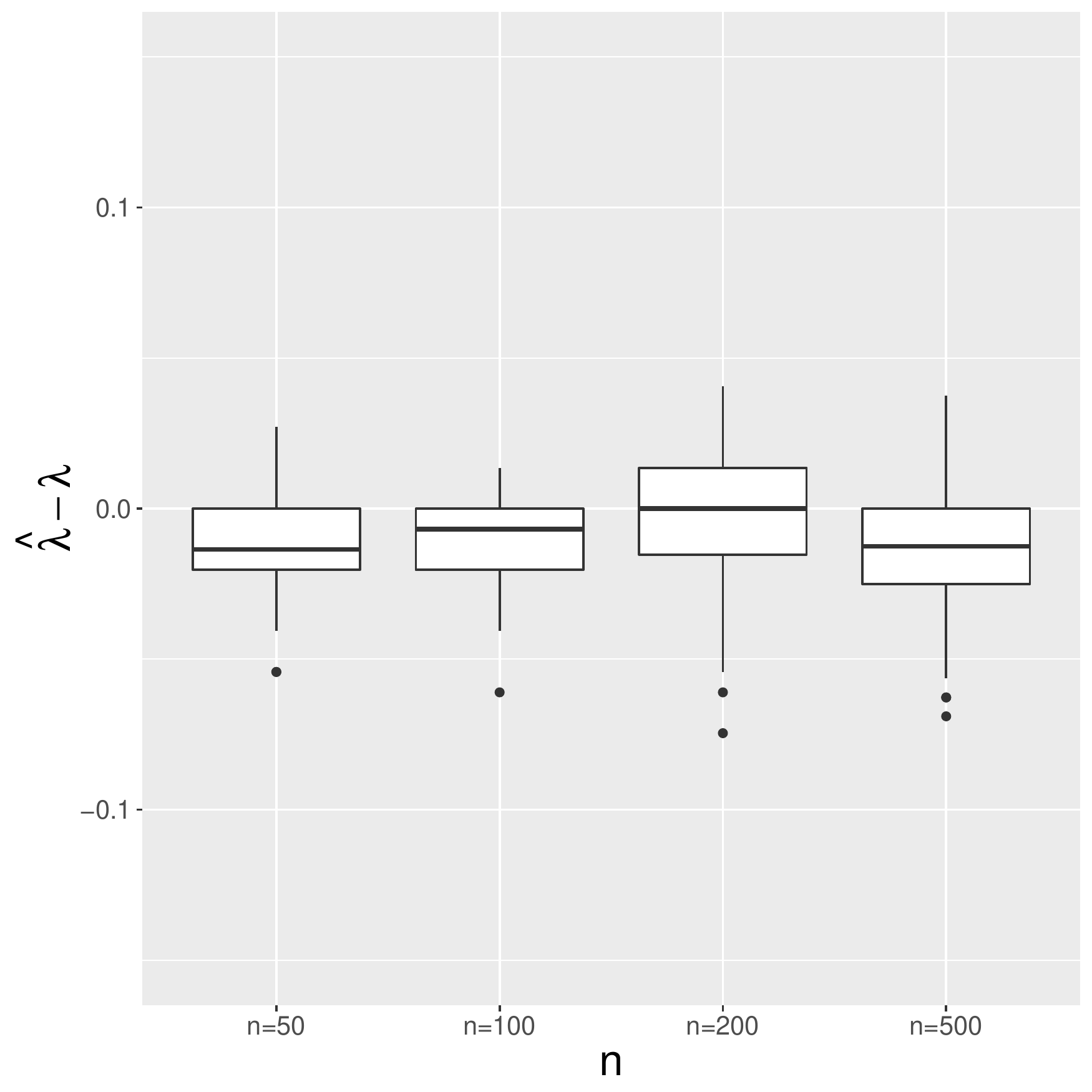}}\\

\end{tabular}
 \caption{A-MSE $\lambda$ selections against the oracle best $\lambda$ for dissimilarities mean square errors.}\label{fags3}
\end{center}
\end{figure}

In Figure \ref{fags3} we present a comparison between the $\lambda$ selection by A-MSE and the $\lambda^\text{oracle}$ that minimizes
\begin{equation}\label{eq:AAG2}
R_{MSE}(\lambda) = \mathbf{E}(\sum_{i>j} (d_{ij}-\hat{d}_{ij}(\lambda))^2).
\end{equation}
We use 60 simulated data sets for each n/p combination with the true models being a power-law distribution as defined in Section \ref{SUP:SEC1}.  Similar results are found when using hubs-based models. A-MSE selections are reasonably close to the best $\lambda$ in almost all presented scenarios, and in all cases, the oracle value of $\lambda$ is within the $95\%$  confidence interval for the median of $\lambda_{AMSE}$.

\newpage
\normalsize
\subsection{AGNES by variable subset approach}
\begin{figure}[ht]
\begin{center}
 \begin{tabular}{ccc}
\subfloat[$\lambda$ errors (p=700) ]{\includegraphics[scale=0.20]{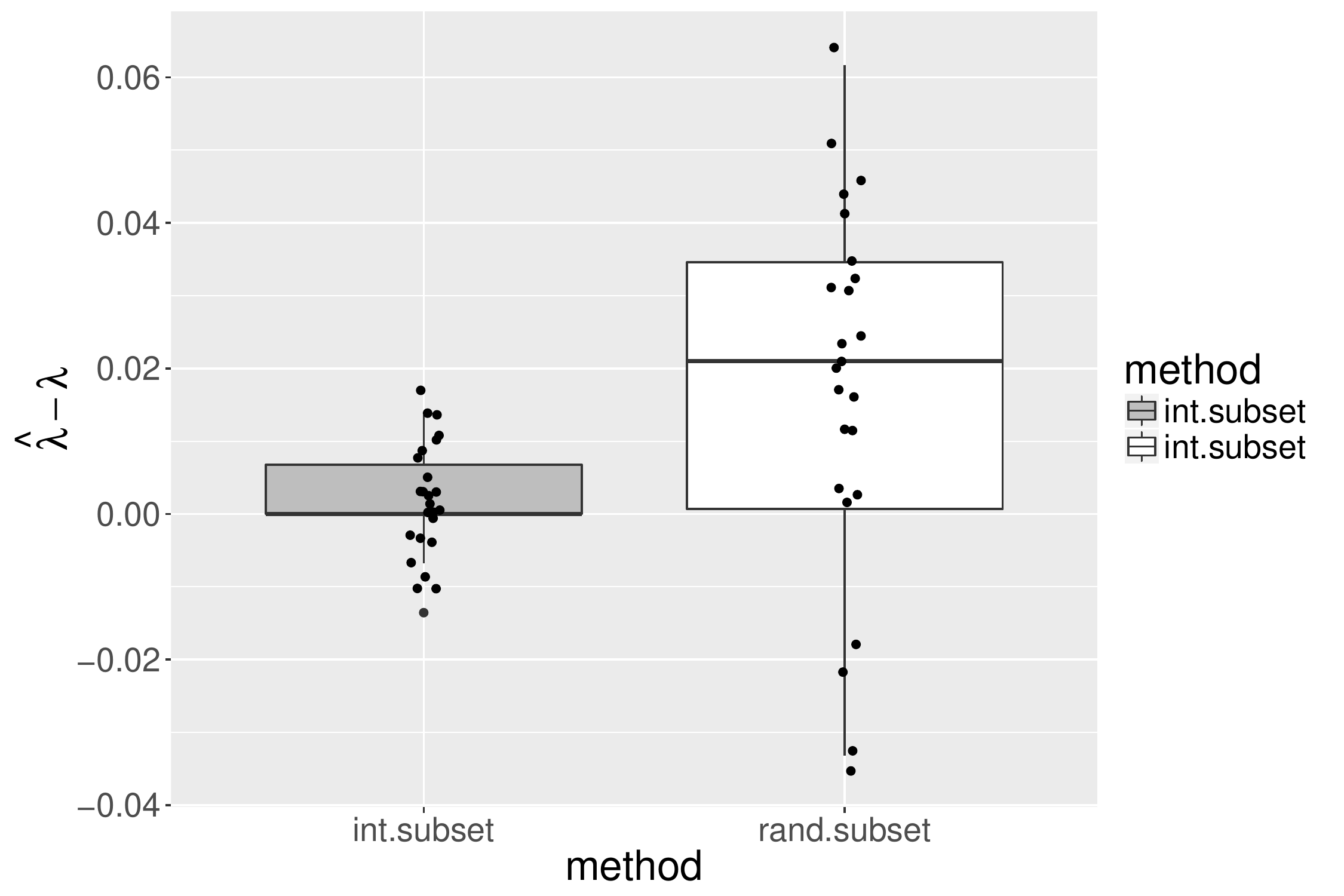}}&
\subfloat[$\lambda$ errors (p=1400) ]{\includegraphics[scale=0.20]{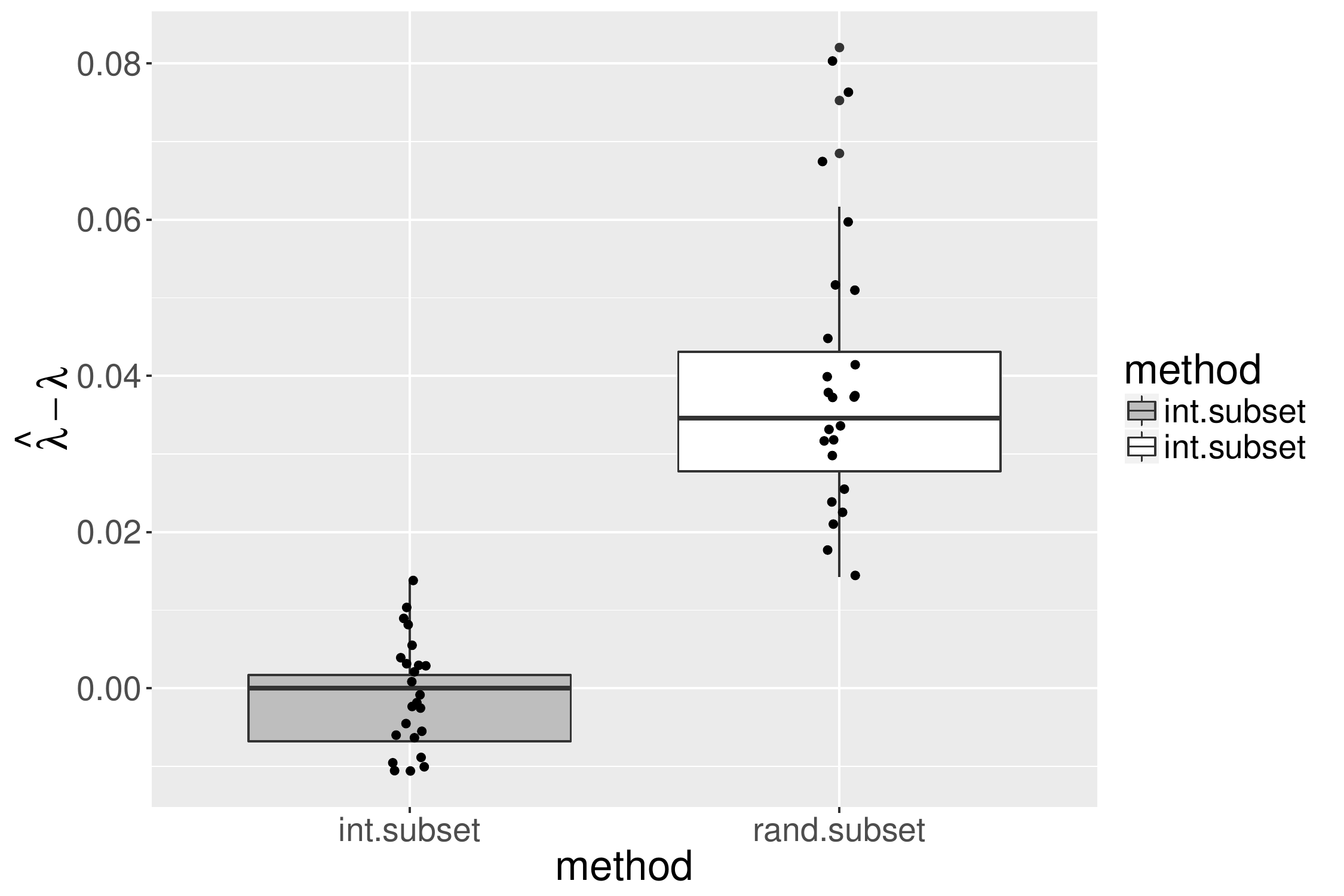}}&
\subfloat[$\lambda$ errors (p=1645) ]{\includegraphics[scale=0.20]{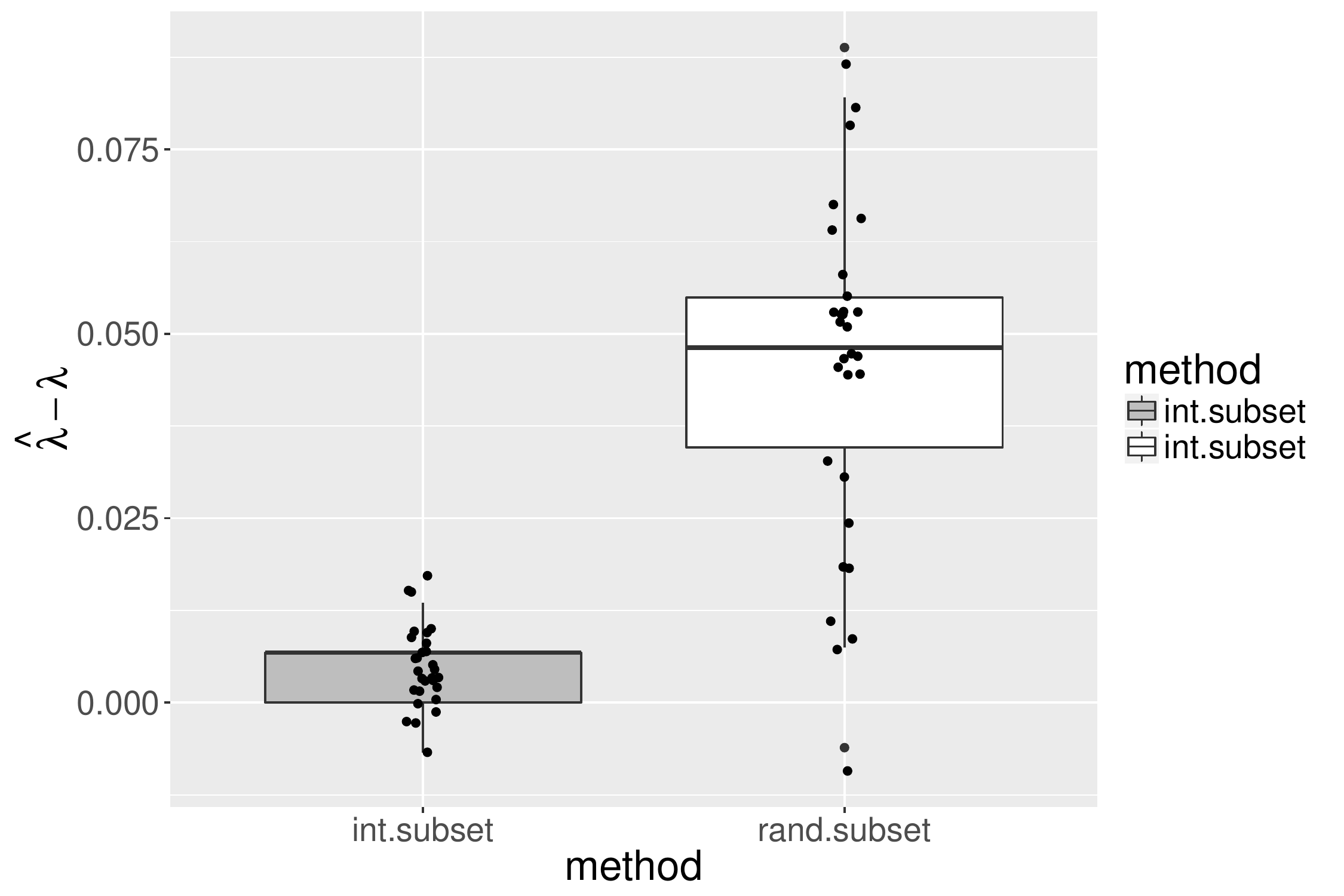}}\\
\subfloat[Relative time reduction (p=700) ]{\includegraphics[scale=0.20]{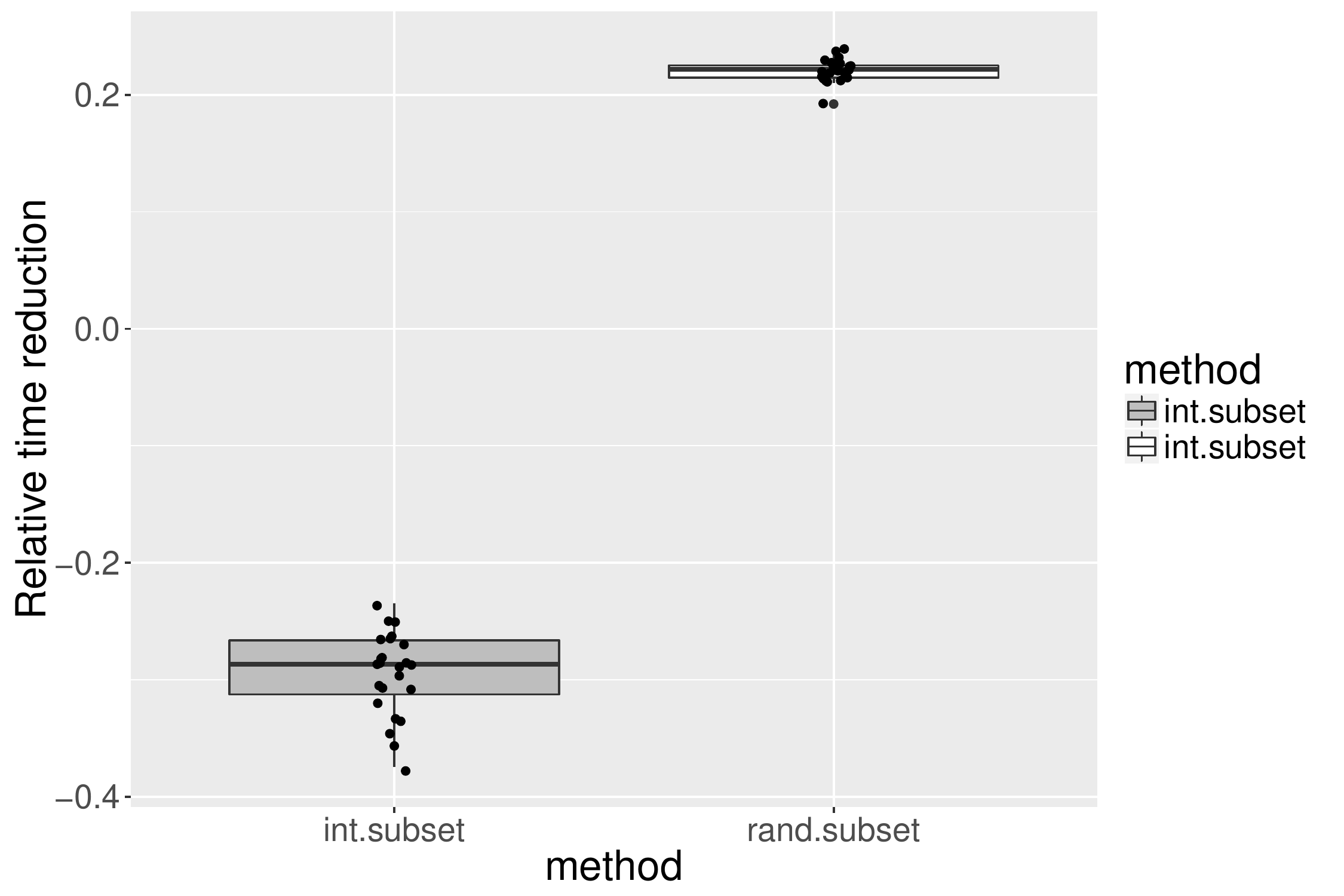}}&
\subfloat[Relative time reduction (p=1400) ]{\includegraphics[scale=0.20]{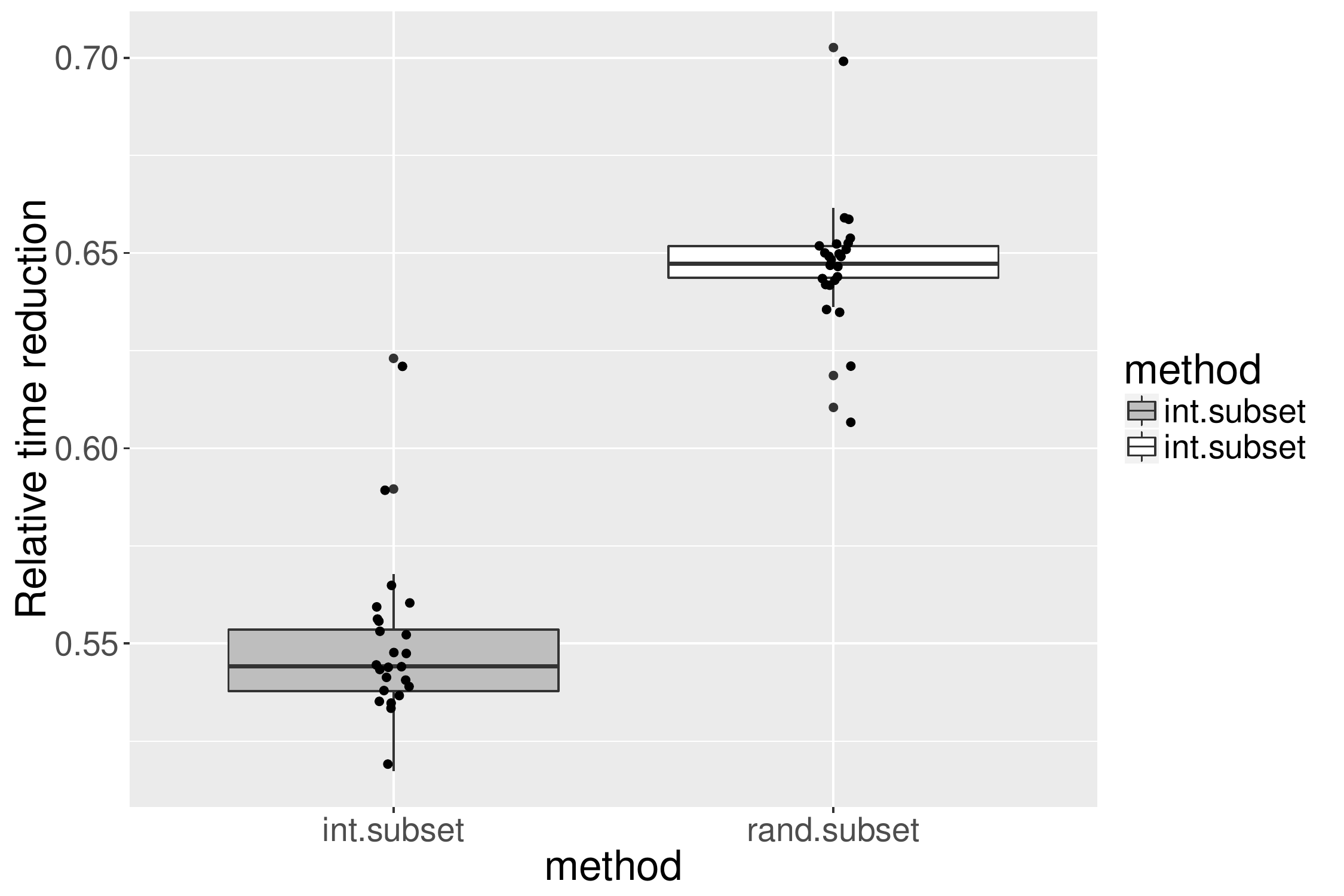}}&
\subfloat[Relative time reduction (p=1645) ]{\includegraphics[scale=0.20]{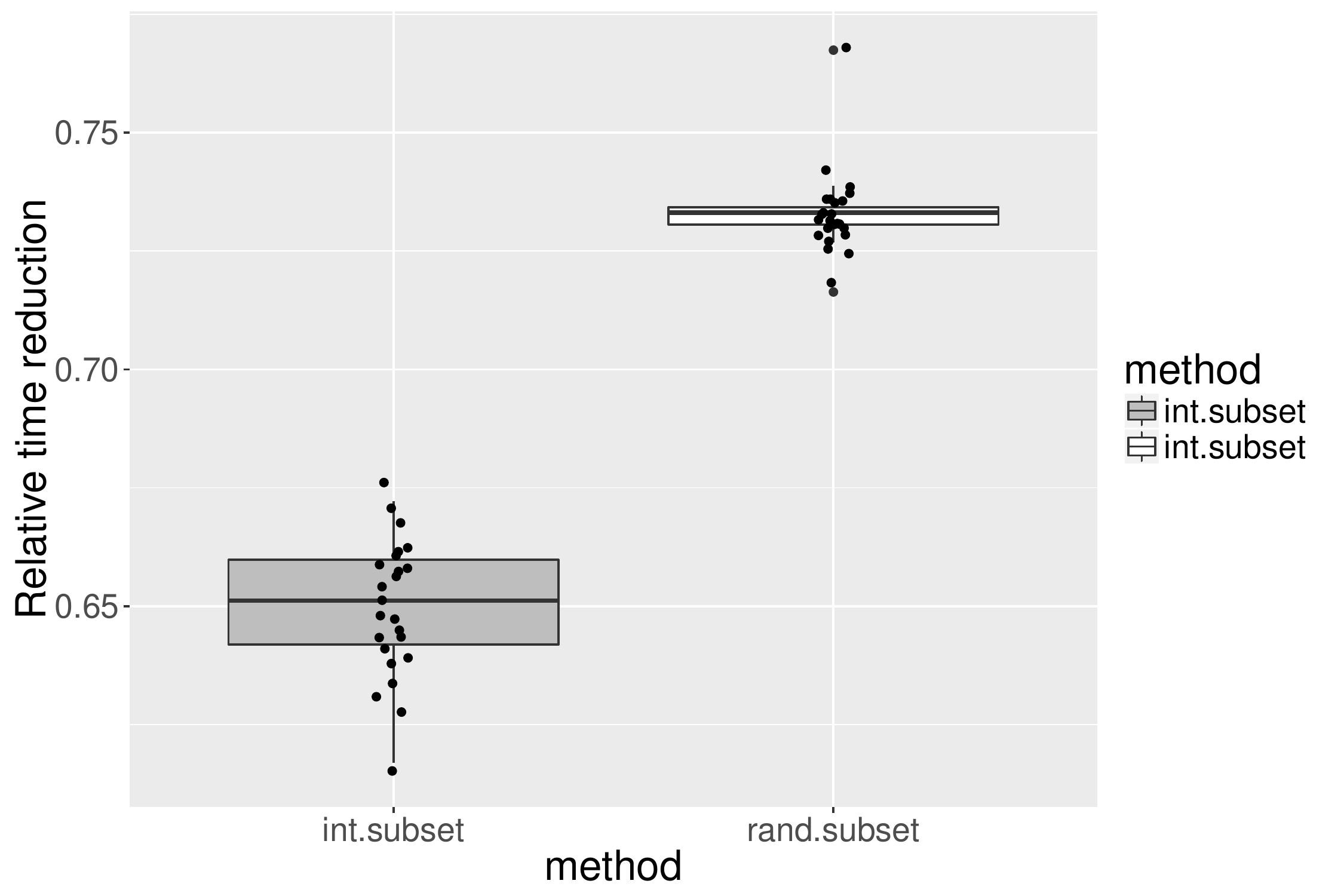}}\\

\end{tabular}
 \caption{Estimation of the best graph by subset selection in the AGNES approach: Errors in the $\lambda$ selection with respect to AGNES with all variables and relative time reduction with respect to the AGNES procedure.}\label{fags2}
\end{center}
\end{figure}

Here, we test two different ways of computing the variable subsets to approximate the AGNES coefficient in high-dimensional data: totally random subsets and the strategy proposed in Section 3.3 in the article (which we denote it by intelligent subsets). In Figure \ref{fags2} we show the $\lambda$ errors (a, b, c) and the relative time reduction (d, e, f) for the subset approaches with respect to the procedure using all the variables. We use 50 replicates for three different power-law true models with large $p$ ($=700, 1400, 1645$) and sample size $n=100$. For large dimension size (i.e. $p>1000$), the subset approach reduces considerably the computational time while maintaining a similar $\lambda$ estimation as the one given using the whole data. 

\footnotesize
\newpage
\bibliography{articlearXiv.bbl}

\end{document}